\documentclass[pra,10pt,twocolumn,superscriptaddress,showpacs]{revtex4-2}
\usepackage{soul}
\usepackage{epsfig}
\usepackage{amsmath}
\usepackage{graphicx}
\usepackage{graphics}
\usepackage{float}
\usepackage{amssymb}
\usepackage{natbib}
\usepackage{epstopdf}
\usepackage{wrapfig}
\usepackage{mathtools}
\usepackage{xcolor}
\RequirePackage{tcolorbox}
\RequirePackage[outline]{contour}
\usepackage{amsfonts}
\usepackage{dsfont}
\usepackage{bm}
\usepackage{physics}
\usepackage{verbatim}
\renewcommand{\bra}[1]{\left\langle #1\right|}
\renewcommand{\ket}[1]{\left| #1\right\rangle}

\newcommand{\opav}[3]{\langle #1 | #2 | #3 \rangle}

\usepackage[english]{babel}

\usepackage{hyperref}
 \hypersetup{
     colorlinks=true,
     linkcolor=blue,
     filecolor=magenta,      
     urlcolor=blue,
     citecolor=blue}
\usepackage{empheq}

\definecolor{fore}{RGB}{249,242,215}
\definecolor{responsecolor}{rgb}{0.1, 0.5, 0.1} 
\definecolor{myblue}{rgb}{.8, .8, 1}
\definecolor{forshading}{RGB}{185,217,255}
\definecolor{bluecomment}{rgb}{0.2, 0.4, 0.7}    
\definecolor{redcomment}{rgb}{0.7, 0.2, 0.2}     
\definecolor{greencomment}{rgb}{0.2, 0.6, 0.2}   
\definecolor{orangecomment}{rgb}{0.9, 0.6, 0.1}  
\newcommand*{\boxedcolor}{blue}
\renewcommand{\boxed}[1]{\textcolor{\boxedcolor}{\fbox{\normalcolor\m@th$\displaystyle#1$}}}
\makeatother

\begin{document}

\title{Sensing high-frequency AC fields via a two-qubit sensor}
\author{Rizwan Abbas}
\affiliation{School of Science \& Engineering, Lahore University of Management Sciences (LUMS), Opposite Sector U, D.H.A, Lahore 54792, Pakistan}
\author{Ali Raza Mirza}
\affiliation{Department of Physics, University of Surrey, GU2 7XH, Guildford, United Kingdom}
\author{Adam Zaman Chaudhry}
\email[]{adam.zaman@lums.edu.pk}
\affiliation{School of Science \& Engineering, Lahore University of Management Sciences (LUMS), Opposite Sector U, D.H.A, Lahore 54792, Pakistan}

\begin{abstract}
    Quantum sensors allow us to measure weak oscillating fields with incredible precision. One common approach is to use the time evolution of a single two-level system (or a qubit) in conjunction with applied control pulses to measure the oscillating field. For high-frequency fields, the time interval required between the applied pulses decreases, meaning that errors due to the finite width of the pulses can become important. This paper presents an alternative scheme that does not rely on applying pulses with short time intervals. Our scheme uses two interacting qubits. In the presence of an oscillating field, the interaction strength changes. The oscillating field can be estimated by measuring the change in this interaction strength. We quantify the precision of this estimate by calculating the Fisher information. We show the effect of noise on our scheme and discuss how control pulses can be applied to mitigate the impact of noise. Importantly, the time interval between these pulses need not be very short. 
\end{abstract}


\maketitle
\section{Introduction} \label{sec1}
The use of quantum systems to measure various physical quantities such as magnetic fields, electric fields, temperature, acceleration, and time has attracted considerable attention recently due to their sensitivity, accuracy, resolution, and size \cite{AslamNatRevPhys2023}. In particular, various schemes utilizing such quantum sensors have been put forward to measure oscillating fields \cite{CappellaroRevModPhys2017,CappellaroPRX2022,Bonizzonninpj2024,SanperaPRA2024}. The ability to measure such fields has various applications in data storage, imaging, and information processing \cite{FreemanScience2001}. One common approach is to use a two-level system, such as a NV center, initially prepared in a superposition state \cite{TaylorNatPhys2008, MazeNature2008, BalasubramanianNature2008, ChangNatNano2008, BalasubramanianNatNano2009, McguinnessNatNano2011, deLangePRL2011, HorowitzPNAS2012, HirosePRA2012, HallMRS2013, HongMRS2013, NusranPRB2013, LoretzPRL2013, LeSageNature2013, GeiselmannNatureNano2013, MagesanPRA2013, CooperNatCommun2014, NusranarXiv2014, DegenRev2014, RondinRev2014, c28, ChaudhryPRA2015}. Due to the field, a phase difference develops as the state of the two-level system evolves, which can then be read out and the field inferred. Since the field is oscillating, it is necessary to apply control pulses. Otherwise, the phase difference does not accumulate due to the repeated reversals of the field \cite{CappellaroRevModPhys2017}. The applied control pulses also have the positive effect of mitigating the effect of noise on the quantum sensor \cite{SuterRev2016}. In fact, analyses have been carried out on the effect of applying such different dynamical decoupling pulse sequences on quantum sensors \cite{PhamPRB2012}. However, measuring high-frequency oscillating fields using quantum sensors is challenging \cite{CappellaroPRX2022}. For example, the time interval between the applied short pulses becomes very short for high-frequency fields, meaning that the usual approximation of treating the pulses via Dirac delta functions can break down. Errors due to the finite width of the pulses then become important \cite{LidarPhysRevAppl2023}. Also, one needs to know the frequency of the field very precisely so that the pulses can be timed correctly. The question we ask is then: can we come up with a scheme to measure high-frequency oscillating fields that does not rely on the application of pulses? As a follow-up to this question, we ask whether, in this scheme, it would still be possible to apply some pulses (albeit not as frequently as before) to mitigate the effect of noise?

To answer these questions, we present an alternative scheme for measuring high-frequency oscillating fields in this paper. Rather than using a single two-level system (or qubit), we propose using two qubits interacting with each other. We show that an oscillating field modifies the interaction strength between the qubits. In fact, in the frame of the oscillating field, we find that the Hamiltonian of the two qubits is, to an excellent approximation, time-independent. By measuring this change in the interaction strength, as is evident from the effective time-independent Hamiltonian, we can deduce the amplitude of the oscillating field. To quantify the precision of our estimate of the oscillating field, we compute the Fisher information and find the measurement that optimizes it. Importantly, we show that our scheme does not rely on knowing the phase of the field. We then show that noise suppresses the buildup of the Fisher information - the longer the decoherence and relaxation timescales, the larger the maximum Fisher information can be. Finally, we demonstrate that we can still apply pulses in our scheme to measure the oscillating field. The pulses considerably lengthen the decoherence and relaxation timescales, thereby significantly increasing the maximum possible Fisher information and hence the precision of our estimate regarding the oscillating field. 

This paper is organized as follows. In Sec.~\ref{sec2} we present our scheme of using two interacting qubits and calculate the Fisher information for the estimation of the oscillating field. In Sec.~\ref{sec3}, we use a simple model to introduce the effect of noise in our scheme and discuss how the Fisher information can be optimized. We also show that pulses can be applied to considerably decrease the impact of noise and thereby increase the maximum Fisher information and the precision of our estimates. Finally, we conclude in Sec.~\ref{sec4}. Two appendices present the technical details of how we model noise in our two-qubit sensor and how measurements can be performed to extract information about the oscillating field. 

\section{The proposed scheme} \label{sec2}
Before introducing our proposed scheme that relies on two qubits, we first briefly recap the usual scheme that relies on a single qubit \cite{CappellaroRevModPhys2017}.

\subsection{Single-qubit sensor}
We begin by assuming that the interaction of the oscillating field to be measured with the single-qubit sensor is described by the Hamiltonian $H(t) = b \cos(\omega t) \sigma_z$
where $b$ is proportional to the amplitude of the field, $\omega$ is its frequency, and $\sigma_z$ is the usual Pauli matrix (note that we will use $\hbar = 1$ throughout). The eigenstates of $\sigma_z$ are denoted as $\ket{0}$ and $\ket{1}$ with eigenvalues $+1$ and $-1$ respectively. The sensor is initially prepared in the state $\ket{\psi(0)} = \frac{1}{\sqrt{2}}(\ket{0} + \ket{1})$. A simple calculation shows that the sensor state at time $t$ is then $\ket{\psi(t)} = \frac{1}{\sqrt{2}}(\ket{0} + e^{i\Phi(t)}\ket{1})$, where the phase is 
\begin{equation}
\label{singlequbitphase}
\Phi(t) = \frac{2b\sin(\omega t)}{\omega}.
\end{equation} 
By measuring this phase $\Phi(t)$, we could measure the amplitude of the oscillating field. However, as is obvious, due to the oscillatory nature of the field, this phase factor does not accumulate as time increases. Having the sensor interact with the field for longer and longer time durations does not help. The usual way around this limitation is to apply control pulses whenever the oscillating field is zero. In particular, once again starting from the initial state $\ket{\psi(0)} = \frac{1}{\sqrt{2}}(\ket{0} + \ket{1})$, we find that $\ket{\psi(t = \frac{\pi}{2\omega})} = \frac{1}{\sqrt{2}}(\ket{0} + e^{i\Phi}\ket{1})$, with $\Phi = \frac{2b}{\omega}$. We now apply a $\pi$ pulse, allow the system to evolve until $t = \frac{3\pi}{2\omega}$, and then apply another $\pi$ pulse. The state is then $\frac{1}{\sqrt{2}}\left(\ket{0} + e^{i\frac{6b}{\omega}}\ket{1}\right)$. Extending the approach to $N$ pulses is straightforward; the key idea is that the phase continues to accumulate due to the applied pulses. However, as mentioned in the introduction, this approach becomes problematic for high-frequency fields due to the very short time interval required between the pulses and the inevitable accumulation of pulse errors.   

\subsection{Two-qubit sensor}
\begin{figure}[h]
    \centering
    \includegraphics[width=0.45\textwidth]{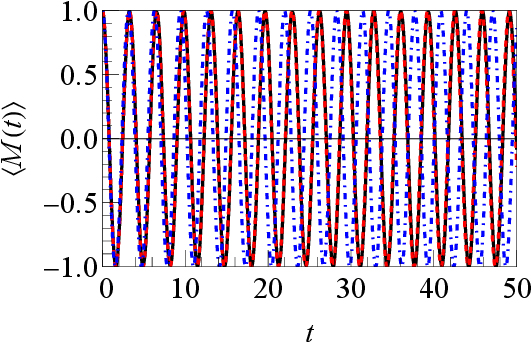}
    \caption{$\langle M(t) \rangle$ as a function of time $t$ with the full time-dependent Hamiltonian (solid, black curve), the effective time-independent Hamiltonian (dashed, red curve), and the original Hamiltonian with no oscillating field present (dot-dashed, blue curve). As usual, we are working in dimensionless units with $\hbar = 1$, and we have set $g = 1$. We have used $b = 1$ and $\omega = 10$.}
    \label{validityofapprox}
\end{figure}

Given the aforementioned fundamental challenges to measure an oscillating field, especially in the high-frequency regime, in this paper, we present an alternative scheme to measure the amplitude of the field. To describe our scheme involving two qubits, we start by writing the Hamiltonian of the two-qubit sensor interacting with the oscillating field as 
\begin{align}
    H(t)=g\sigma^{(1)}_x \sigma^{(2)}_x + b \cos(\omega t) [\sigma^{(1)}_z +\sigma^{(2)}_z ].
    \label{fullH}
\end{align}
Here, the superscripts label the qubits, $\sigma_x$ and $\sigma_z$ are the usual Pauli matrices, and $g$ is now the interaction strength between the qubits. As before, $b$ is proportional to the amplitude of the field, and $\omega$ is its frequency. We first transform to the frame of the oscillating field \cite{c6, HudaibaEJPD2022, IrfanPRA2024}. This is done via the unitary operator
\begin{align*}
    U_b(t)
    &=\exp[-ib \left(\sigma^{(1)}_z +\sigma^{(2)}_z \right) C(t)],
\end{align*}
with $C(t) = \sin(\omega t)/\omega$. In this frame, the Hamiltonian becomes, $$\widetilde{H}(t)=U^{\dagger}_b(t)g\sigma^{(1)}_x \sigma^{(2)}_x U_b(t).$$ To make further progress, we introduce the raising and lowering operators as $\sigma_\pm = \frac{1}{2}(\sigma_x \pm i\sigma_y)$. We then find $\sigma^{(1)}_x \sigma^{(2)}_x = 
\sigma^{(1)}_{+} \sigma^{(2)}_{+} +
\sigma^{(1)}_{-} \sigma^{(2)}_{-}+\sigma^{(1)}_{+} \sigma^{(2)}_{-}+\sigma^{(1)}_{-} \sigma^{(2)}_{+}]$. Using the identity 
 $e^{ibC(t)\sigma_z}\sigma_{\pm} e^{-ibC(t)\sigma_z}=\sigma_{\pm}e^{\pm 2ibC(t)}$, we get
\begin{align}
    \widetilde{H}(t)
    =g\Big[
    e^{4ibC(t)}\sigma^{(1)}_{+} \sigma^{(2)}_{+} 
    &+e^{-4ibC(t)}\sigma^{(1)}_{-} \sigma^{(2)}_{-}\notag
\\
    &+\sigma^{(1)}_{+} \sigma^{(2)}_{-} +
    \sigma^{(1)}_{-} \sigma^{(2)}_{+}\Big].
    \label{Hfieldframe}
\end{align}
Note that the modification to the Hamiltonian due to the oscillating field only comes about because the coupling between the qubits does not commute with the coupling of the qubits with the oscillating field. Our calculations are exact so far and valid for any applied field frequency. Now, according to the Jacobi-Anger expansion, $\text{exp}\left[4ib\sin(\omega t)/\omega\right] = \sum_{n=-\infty}^{n=+\infty}J_n\left(\frac{\pm4b}{\omega} \right)e^{in\omega t}$,  
    where $J_n(x)$ is the Bessel function of the first kind. For large $\omega$ such that $\omega \gg g$, we can expect that the high-frequency terms in this expansion lead to time evolution that averages out to zero. As such, for high-frequency fields, the Hamiltonian effectively becomes    
    \begin{align*}
     \widetilde{H}
    =g\Big[
    A (\sigma^{(1)}_{+} \sigma^{(2)}_{+} +
    \sigma^{(1)}_{-} \sigma^{(2)}_{-})+\sigma^{(1)}_{+} \sigma^{(2)}_{-} +
    \sigma^{(1)}_{-} \sigma^{(2)}_{+}\Big],
\end{align*}
where $A = J_0\left(\frac{4b}{\omega}\right)$. This can be written more transparently as 
\begin{align} \label{a3}
    \widetilde{H}
    =\frac{g}{2}\left[(1 + A)\sigma^{(1)}_{x}\sigma^{(2)}_{x} 
    +(1 - A)\sigma^{(1)}_{y}\sigma^{(2)}_{y}\right].
\end{align}
Notice that this effective Hamiltonian is time-independent. The eigenvalues of this Hamiltonian are $\lambda_1 = Ag$, $\lambda_2 = -Ag$, $\lambda_3 = g$, and $\lambda_4 = -g$, with the corresponding eigenstates $\ket{\lambda_1} = \frac{1}{\sqrt{2}}\left(\ket{00} + \ket{11}\right)$, $\ket{\lambda_2} = \frac{1}{\sqrt{2}}\left(\ket{00} - \ket{11}\right)$, $\ket{\lambda_3} = \frac{1}{\sqrt{2}}\left(\ket{01} + \ket{10}\right)$, and $\ket{\lambda_4} = \frac{1}{\sqrt{2}}\left(\ket{01} - \ket{10}\right)$ respectively. Recall that the states $\ket{0}$ and $\ket{1}$ are eigenstates $\sigma_z$ with $\sigma_z\ket{0} = \ket{0}$ and $\sigma_z\ket{1} = -\ket{1}$. These eigenstates then motivate us to consider the initial two-qubit state as $\ket{\psi(0)} = \ket{00} = \frac{1}{\sqrt{2}}\left(\ket{\lambda_1} + \ket{\lambda_2}\right)$. Then, at time $t$, $\ket{\psi(t)} = \frac{1}{\sqrt{2}}\left(\ket{\lambda_1} + e^{i2Ag t}\ket{\lambda_2}\right)$. To extract the value of $A$ (and hence the value of $b$), 
let us define the population-difference observable as
$M=\ket{00}\bra{00}-\ket{11}\bra{11}$. If we measure this observable at time $t$, the probability of getting measurement result equal to $+1$ is $p_{+1} = \cos^2(Ag t)$, while the probability of getting measurement result $-1$ is obviously $p_{-1} = \sin^2(Ag t)$. Consequently, $\langle M(t) \rangle = p_{+1} - p_{-1} = \cos(2Agt)$. Thus, by measuring the expectation value of the observable $M$, we can deduce the value of $A$ and hence the value of $b$.
In Fig.~\ref{validityofapprox}, we show $\langle M(t) \rangle$ using our effective Hamiltonian approach (the dot-dashed red curve). In contrast, the solid, black curve shows the dynamics by numerically solving the full time-dependent Hamiltonian given in Eq.~\eqref{fullH}. Also, the dotted-dashed blue curve shows the dynamics without the oscillating field. There are two important points to note here. First, the fact that the solid, black curve overlaps with the dashed, red curve illustrates the validity of our effective time-independent Hamiltonian approach. Second, the dot-dashed blue curve is precisely what underlies our scheme. The presence of the oscillating field affects the coupling between the qubits and therefore affects $\langle M(t) \rangle$. 

Before moving on, let us note that, given $\ket{\psi(t)} = \frac{1}{\sqrt{2}}\left(\ket{\lambda_1} + e^{i2Ag t}\ket{\lambda_2}\right)$, we find that $\opav{\psi(t)}{(\ket{0}\bra{0} \otimes \mathds{1})}{\psi(t)} = \cos^2(Agt) = p_{+1}$. Similarly, $\opav{\psi(t)}{(\ket{1}\bra{1} \otimes \mathds{1})}{\psi(t)} = \sin^2(Agt) = p_{-1}$. In other words, to find $\langle M(t) \rangle$, we can measure only one of the qubits. How this measurement is performed in practice depends, of course, on the physical realization of the qubits. For example, if we realize our qubits using trapped ions, the state of a qubit can be read out using ion fluorescence \cite{MyersonPRL2008}.

\begin{figure}[b!]
    \centering
    \includegraphics[width=0.45\textwidth]{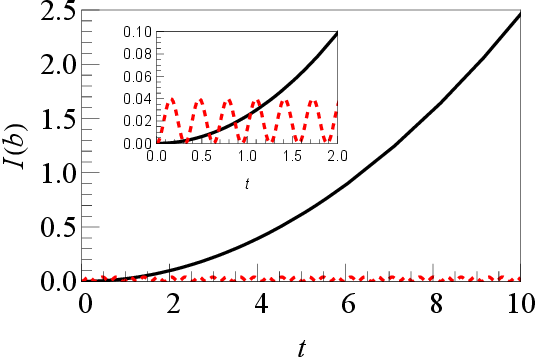}
    \caption{Plot of the Fisher information $I(b)$ as a function of time $t$. The solid, black curve is the Fisher information with our two-qubit sensor, while the dashed, red curve is the Fisher information with a single qubit sensor. As usual, we are working in dimensionless units with $\hbar = 1$, and we have set $g = 1$. Also, $b = 1$ and $\omega = 10$. The figure in the inset is the same as the main figure, except that we have zoomed in to show the behavior of the Fisher information at small times.}
    \label{fishercomparison}
    \end{figure}

The precision of our estimate can be quantified by the Fisher information \cite{Jacobsbook}. The probabilities can be written as $p(n|b)$, with $n = \pm 1$, showing that they depend on the parameter $b$. The Fisher information for $b$ can then be computed as 
\begin{equation}   
I(b) = \sum_n p(n|b) \left(\frac{\partial}{\partial b} \ln p(n|b) \right)^2. 
\end{equation}
Using the form of $p(n|b)$, we find that 
$$I(b) = (g t)^2 \sin^2(2Ag t) \left(\frac{\partial A}{\partial b}\right)^2 \sum_n \frac{1}{p(n|b)}, $$
which leads to 
$$I(b) = (2g t)^2 \left(\frac{\partial A}{\partial b}\right)^2. $$
Since $A = J_0\left(\frac{4b}{\omega}\right)$, $\left(\frac{\partial A}{\partial b}\right) = \frac{4}{\omega} J_1\left(\frac{4b}{\omega}\right)$. Thus,
\begin{equation}
\label{classicalFisherorig}
I(b) = \left(\frac{8g t}{\omega}\right)^2 \left[J_1\left(\frac{4b}{\omega}\right)\right]^2.
\end{equation}
This is a key result. A longer interaction time $t$ and a stronger qubit-qubit interaction strength $g$ rapidly improve the Fisher information and thus the precision of our estimate of $b$. It is instructive to compare this Fisher information for our two-qubit sensor with the Fisher information obtained using a single-qubit sensor in the absence of pulses. For the latter, we estimate $b$ using the phase $\Phi(t)$ given in Eq.~\eqref{singlequbitphase}. This leads to the Fisher information $I(b) = \frac{4}{\omega^2}\sin^2(\omega t)$ \cite{c28}. This is clearly very different from the Fisher information for our two-qubit sensor. Fig.~\ref{fishercomparison} clearly shows the drastic increase in Fisher information with our two-qubit sensor compared to the single-qubit sensor.

A natural question to ask at this point is: can we do better? That is, can we perform some other measurement (instead of measuring the observable $M$) that will allow us to estimate $b$ with greater precision? To answer this question, we compute the quantum Fisher information (QFI) since this optimizes over the measurement performed \cite{Jacobsbook,BenedettiPRA2018,ChaudhryPRA2021,ChaudhrySciRep2024,mirza2024impact}. Given the density matrix $\rho(t)$ for the sensor, the QFI can be computed via 
\begin{equation}
I_Q(b) = 2 \sum_{k,l} \frac{|\opav{\rho_k}{\frac{\partial \rho}{\partial b}}{\rho_l}|^2}{\rho_k + \rho_l}, 
\end{equation}
where $\ket{\rho_k}$ is the $k^\text{th}$ eigenstates of the density matrix $\rho(t)$ with corresponding eigenvalue $\rho_k$. The sum is over all $k$ and $l$ such that $\rho_k + \rho_l \neq 0$. Now, given that the sensor state $\rho(t) = \ket{\psi(t)}\bra{\psi(t)}$ that we are considering is pure, and given the form of $\ket{\psi(t)}$, we find that
$$ I_Q(b) = 4 |\opav{\rho_2}{\frac{\partial \rho}{\partial b}}{\rho_1}|^2, $$
where $\ket{\rho_1} = \cos(Ag t)\ket{00} - i \sin(Ag t)\ket{11}$ and $\ket{\rho_2} = \sin(Ag t)\ket{00} + i \cos(Ag t)\ket{11}$. After further simplification, we get 
$$I_Q(b) = \left(2g t \frac{\partial A}{\partial b}\right)^2 = \left(\frac{8g t}{\omega}\right)^2 \left[J_1 \left(\frac{4b}{\omega}\right)\right]^2. $$
This is exactly the same as the Fisher information we found for measuring the observable $M$ [see Eq.~\eqref{classicalFisherorig}]. Therefore, we are performing the optimal measurement to estimate $b$. 

\begin{figure}[t!]
    \centering
    \includegraphics[width=0.45\textwidth]{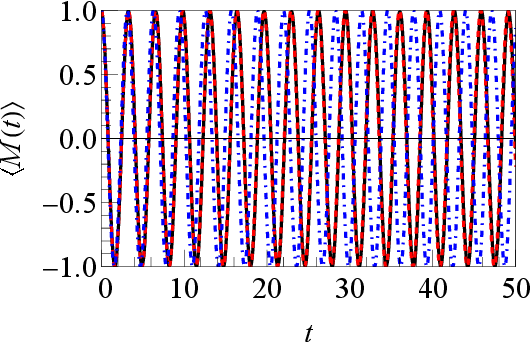}
    \caption{$\langle M(t) \rangle$ as a function of time $t$ with the full time-dependent Hamiltonian with $\phi = 30^\circ$ (solid, black curve), the effective time-independent Hamiltonian in Eq.~\eqref{a3} (dashed, red curve), and the original Hamiltonian with no oscillating field present (dot-dashed, blue curve). As usual, we are working in dimensionless units with $\hbar = 1$, and we have set $g = 1$. We again have $b = 1$ and $\omega = 10$.}
    \label{validityofapproxwithphase}
\end{figure}

Let us now generalize our treatment to the scenario where the oscillating field has an unknown phase. In other words, the time-dependence of this field is given by $\cos(\omega t + \phi)$ rather than simply $\cos(\omega t)$. Here $\phi$ is the unknown phase. If we use a single qubit as the sensor, timing the pulses would be problematic since we do not know when the field flips. The usual way around this is to use different pulse sequences \cite{c28}, and use the results we obtain using different pulse sequences. As such, using pulses to estimate the amplitude of a high-frequency field with an unknown phase becomes even more demanding. In contrast, as we now show, our scheme can naturally cater to this scenario of unknown phase - no additional effort is required. The Hamiltonian that describes the interaction of our two-qubit sensor with the oscillating field is now 
\begin{align}
    H(t)=g\sigma^{(1)}_x \sigma^{(2)}_x + b (\cos\omega t + \phi) [\sigma^{(1)}_z +\sigma^{(2)}_z ].
    \label{fullHwithphase}
\end{align} 
As before, we transform to the frame of the oscillating field. The resulting Hamiltonian is of the same form as before [see Eq.~\eqref{Hfieldframe}], but now $C(t) = \frac{\sin(\omega t + \phi) - \sin \phi}{\omega}$. Making the high-frequency approximation that we made before, we can now write   
\begin{align*}
    \widetilde{H}(t)
    =g\Big[
    A_c A_s e^{-i\Phi} \sigma^{(1)}_{+} \sigma^{(2)}_{+} 
    &+ A_c A_s e^{i\Phi}\sigma^{(1)}_{-} \sigma^{(2)}_{-}\notag
\\
    &+\sigma^{(1)}_{+} \sigma^{(2)}_{-} +
    \sigma^{(1)}_{-} \sigma^{(2)}_{+}\Big],
\end{align*}
where $A_c = J_0 \left(\frac{\pm  4b\cos\phi}{\omega} \right)$, $A_s = J_0\left(\frac{\pm  4b\sin\phi}{\omega} \right)$, and $\Phi = 4b\sin\phi / \omega$. Since the arguments of the Bessel functions are small, we have $A_c A_s \approx \left[ 1 - \frac{1}{2}\left(\frac{4b}{\omega}\cos \phi\right)^2\right]\left[ 1 - \frac{1}{2}\left(\frac{4b}{\omega}\sin \phi\right)^2\right] \approx 1 - \frac{1}{2}\left(\frac{4b}{\omega}\right)^2 \approx J_0\left(\frac{4b}{\omega}\right)$. Therefore, for the high-frequency fields that we are dealing with, the Hamiltonian is effectively 
\begin{align}
\label{Hfieldwithphase}
    \widetilde{H}
    =g\Big[
    A e^{-i\Phi} \sigma^{(1)}_{+} \sigma^{(2)}_{+} 
    &+ A e^{i\Phi}\sigma^{(1)}_{-} \sigma^{(2)}_{-}\notag
\\
    &+\sigma^{(1)}_{+} \sigma^{(2)}_{-} +
    \sigma^{(1)}_{-} \sigma^{(2)}_{+}\Big].
\end{align}
The parameter $A$ in the Hamiltonian has picked up a phase. The eigenstates of this Hamiltonian are $\ket{\lambda_1} = \frac{1}{\sqrt{2}}\left(\ket{00} + e^{i\Phi}\ket{11}\right)$, $\ket{\lambda_2} = \frac{1}{\sqrt{2}}\left(\ket{00} - e^{i\Phi}\ket{11}\right)$, $\ket{\lambda_3} = \frac{1}{\sqrt{2}}\left(\ket{01} + e^{i\Phi}\ket{10}\right)$, and $\ket{\lambda_4} = \frac{1}{\sqrt{2}}\left(\ket{01} - e^{i\Phi}\ket{10}\right)$, with corresponding eigenvalues $\lambda_1 = Ag$, $\lambda_2 = -Ag$, $\lambda_3 = g$, and $\lambda_4 = -g$. As before, the initial state that we consider is $\ket{\psi(0)} = \ket{00}$. Then, at time $t$, $\ket{\psi(t)} = \frac{1}{\sqrt{2}}\left(\ket{\lambda_1} + e^{i2Ag t}\ket{\lambda_2}\right)$. We again measure the observable $M$. The probabilities of getting measurement results $+1$ and $-1$ are $\cos^2(Ag t)$ and $\sin^2(Ag t)$ respectively. These are the same as we found previously. In other words, our scheme works for any non-zero value of $\phi$; if we start from the state $\ket{00}$, the dynamics are the same as those obtained via the effective Hamiltonian in Eq.~\eqref{a3}. This is illustrated in Fig.~\ref{validityofapproxwithphase}. The overlap of the dashed, red curve (obtained by using $\phi = 30^\circ$) and the solid, black curve [obtained by using the effective time-independent Hamiltonian in Eq.~\eqref{a3}] illustrates that in our high-frequency regime, $\langle M(t) \rangle$ is independent of the phase $\phi$.

\section{Introducing noise} \label{sec3}

Until now, we have ignored the effect of noise in our scheme for estimating the field amplitude. In the absence of noise, we have concluded that the Fisher information for the estimate will keep increasing with time. In reality, due to the inevitable presence of noise, the Fisher information will decrease at long times as relaxation and dephasing kick in. To model the effect of noise, we recall that $p_{+1} = \frac{1}{2}\left[1 + \cos(2A g t)\right]$ and $p_{-1} = \frac{1}{2}\left[1 - \cos(2A g t)\right]$. Due to dephasing, the phase gets scrambled; to take this into account, we modify these expressions to $p_{+1} = \frac{1}{2}\left[1 + f_2(t)\cos(2A g t)\right]$ and $p_{-1} = \frac{1}{2}\left[1 - f_2(t)\cos(2A g t)\right]$, where the function $f_2(t) = e^{-t/T_2}$ decreases with time and approaches zero at long times. Here $T_2$ is the dephasing timescale. We also consider amplitude damping for the two qubits (see Appendix A for details). The probabilities then become $p_{+1} = \frac{f_1(t)}{2}\left[1 + f_2(t)\cos(2A g t)\right]$, $p_{-1} = \frac{f_1(t)}{2}\left[1 - f_2(t)\cos(2A g t)\right]$, and $p_0 = 1 - p_{+1} - p_{-1}$. Here $f_1(t) = \frac{1}{2}(1 + e^{-t/T_1})$, with $T_1$ the relaxation timescale (we refer to $T_1$ as the relaxation timescale since $f_1(t)$ comes from amplitude damping). With these probabilities, we now calculate the Fisher information for our estimate of $b$. Defining $F(t) = f_1(t)f_2(t)$, we first note that $\left(\frac{\partial p_{+1}}{\partial b}\right)^2  = \left(\frac{4gt}{\omega}\right)^2 [F(t)]^2 \sin^2(2Ag t) \left[J_1\left(\frac{4b}{\omega}\right)\right]^2 = \left(\frac{\partial p_{-1}}{\partial b}\right)^2$. Moreover, $\frac{\partial p_0}{\partial b} = 0$. We then find the Fisher information to be 
\begin{equation}
\label{fishereqwithnoise}
I(b) = \left(\frac{8g t}{\omega}\right)^2 \left[J_1 \left(\frac{4b}{\omega}\right)\right]^2 \mathcal{F}(b,t), 
\end{equation}
where 
\begin{equation}
\mathcal{F}(b,t) = \left[F(t)\right]^2 \frac{\sin^2(2Ag t)}{f_1(t)[1 - [f_2(t)]^2 \cos^2(2Ag t)]}.
\end{equation}
\begin{figure}[b!]
    \centering
    \includegraphics[width=0.45\textwidth]{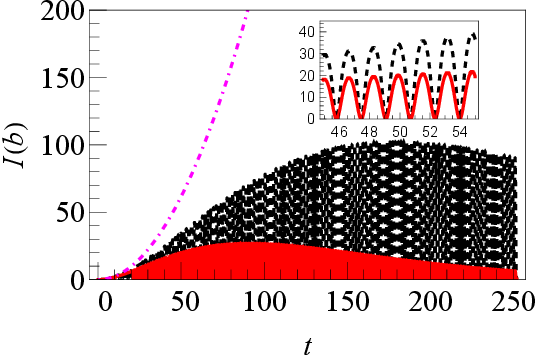}
    \caption{$I(b)$ as a function of time $t$. Here we are using $f_1(t) = \frac{1}{2}\left(1 + e^{-t/T_1}\right)$ and $f_2(t) = e^{-t/T_2}$. The dashed, black curve shows the behavior of the Fisher information with $T_1 = 300$ and $T_2 = 200$, while the solid, red curve shows the behavior with $T_1 = 200$ and $T_2 = 100$. The dot-dashed magenta curve is the Fisher information in the absence of noise. As usual, we are working in dimensionless units with $\hbar = 1$, and we have set $g = 1$. Also, $b = 1$ and $\omega = 10$. The figure in the inset is the same as the main figure, except that we have zoomed in to show the smooth oscillations of the Fisher information [see Eq.~\eqref{fishereqwithnoise}].}
    \label{fisherwithnoise}
    \end{figure}
Note that $\mathcal{F}(b,t)$ takes into account the noise. In the absence of noise, it is easy to verify that $\mathcal{F}(b,t) = 1$. If we have only dephasing so that $f_1(t) = 1$, then we  get $\mathcal{F}(b,t) = [f_2(t)]^2 \frac{\sin^2(2Ag t)}{1 - [f_2(t)]^2 \cos^2(2Ag t)}$. These expressions imply that in the presence of noise, the behavior of the Fisher information is not as simple as first increasing and then decreasing. In fact, whenever $2Ag t$ is an integer multiple of $\pi$, the Fisher information is zero. This behavior is illustrated in Fig.~\ref{fisherwithnoise}, where it can be seen that the Fisher information repeatedly becomes zero in the presence of noise. Two further important points can be made regarding this figure. First, the Fisher information in the presence of noise is obviously bounded - it no longer keeps increasing. This maximum value depends on how strong the influence of the noise is. Smaller values of $T_1$ and $T_2$ decrease the maximum value of the Fisher information. Second, the time at which this maximum value is obtained also depends on the noise. With smaller values of $T_1$ and $T_2$, the optimal value of time decreases. It is also worth mentioning that to find $p_{+1}$, for example, we can no longer measure only a single qubit. Due to the noise, the state of the two qubits is no longer restricted to the subspace spanned by $\ket{00}$ and $\ket{11}$. We now need to perform a simultaneous measurement of both the qubits. Such measurements have been performed - see, for example, Ref.~\cite{ZhukasPRA2021} where such measurements are performed for a trapped ion system. Moreover, as we argue in Appendix B, the measurement on the two qubits need not be exactly simultaneous.

\begin{figure}[h]
    \centering
    \includegraphics[width=0.45\textwidth]{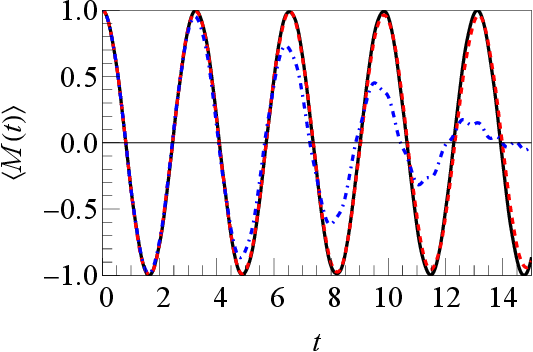}
    \caption{$\langle M(t) \rangle$ (in the toggling frame) as a function of time $t$ without noise (solid, black curve), with noise (dot-dashed, blue curve), and with noise but with pulses also applied (red, dashed curve). For the noise (see the main text), we have used $\mu = 0$, $\sigma = 0.2$, and $\tau = 50$. We have taken an average of $50$ simulations. As usual, we are using dimensionless units with $\hbar = 1$, and $g = 1$, $b = 1$, and $\omega = 10$. }
    \label{labelnoiseandpulses}
\end{figure}

To reduce the effect of noise, we can apply control pulses. As mentioned before, pulses are applied whenever the field is zero when measuring an oscillating field with a single qubit. These pulses not only lead to an accumulation of the phase difference but also suppress the effect of noise. As we show below, pulses can also be applied in our scheme to reduce the impact of noise. These pulses need not be applied very rapidly; however, the time interval between the pulses needs to be shorter than the noise correlation time so that the noise can effectively be suppressed. Similar to the XY4 pulse sequence (but applied to both qubits) \cite{LidarPhysRevAppl2023}, we consider that at times $\tau_1$, $\tau_2$, $\hdots$, $\tau_M$ we apply the unitary operator $e^{-i\pi \sigma_z^{(1)}/2} e^{-i\pi \sigma_z^{(2)}/2}$ (hereafter referred to as `Z pulses'), while at times $t_1$, $t_2$, $\hdots$, $t_N$ we apply the unitary operator $e^{-i\pi \sigma_x^{(1)}/2} e^{-i\pi \sigma_x^{(2)}/2}$ (hereafter referred to as `X pulses'). Such a pulse sequence can effectively suppress the effect of noise of the form $B_{11}\sigma_x^{(1)} + B_{12}\sigma_x^{(2)} + B_{21}\sigma_y^{(1)} + B_{22}\sigma_y^{(2)} + B_{31}\sigma_z^{(1)} + B_{32}\sigma_z^{(2)}$, where the $B_{ij}$ are arbitrary environment operators or time-dependent functions describing classical noise. Our objective is to show that our scheme of estimating the oscillating field works in the presence of these applied pulses. To show this, we start from the Hamiltonian 
\begin{equation}
    H(t)=g\sigma^{(1)}_x \sigma^{(2)}_x +b \cos(\omega t) [\sigma^{(1)}_z +\sigma^{(2)}_z ] + H_p(t),
\end{equation}
where $H_p(t)$ is the Hamiltonian that describes the application of the pulses. We first transform to the frame of the $Z$ pulses, and then to the frame of the $X$ pulses. This leads to the Hamiltonian (in the `toggling frame') 
$$ H_{\text{tf}} = g\sigma^{(1)}_x \sigma^{(2)}_x + s(t) b \cos(\omega t) [\sigma^{(1)}_z +\sigma^{(2)}_z ]. $$
Here $s(t) = \pm 1$ depending on the time. In particular, $s(t) = 1$ for $0 \leq t \leq t_1$, $s(t) = -1$ for $t_1 \leq t \leq t_2$, 
$s(t) = 1$ for $t_2 \leq t \leq t_3$, and so on. The form of this Hamiltonian gives us hope that the change in the interaction is preserved in the presence of the applied pulses. Notice that $A = J_0\left(\frac{4b}{\omega}\right) = J_0\left(-\frac{4b}{\omega}\right)$, meaning that $A$ does not change if $b$ flips sign. To show this clearly, we make a further transformation to the frame of the oscillating field. This transformation is carried out by the unitary operator 
$$U_b(t) = e^{-ib(\sigma_z^{(1)} + \sigma_z^{(2)})\int_0^t dt' \, s(t') \cos(\omega t')}. $$
The transformed Hamiltonian is then of the same form as Eq.~\eqref{Hfieldframe}, namely  
 \begin{align*}
    \widetilde{H}(t)
    =g\Big[
    e^{4ibC(t)}\sigma^{(1)}_{+} \sigma^{(2)}_{+} 
    &+e^{-4ibC(t)}\sigma^{(1)}_{-} \sigma^{(2)}_{-}\notag
\\
    &+\sigma^{(1)}_{+} \sigma^{(2)}_{-} +
    \sigma^{(1)}_{-} \sigma^{(2)}_{+}\Big],
\end{align*}
except that now $C(t) = \int_0^t dt' \, s(t') \cos(\omega t')$. This can be worked out to be $C(t) = \int_0^{t_1} dt' \cos(\omega t') - \int_{t_1}^{t_2} dt' \cos(\omega t') + \hdots + (-1)^N \int_{t_N}^t dt' \cos(\omega t') = 2 \sum_{n \in \text{odd}} \frac{\sin(\omega t_n)}{\omega} - 2 \sum_{n \in \text{even}} \frac{\sin(\omega t_n)}{\omega} + (-1)^N \frac{\sin(\omega t)}{\omega}$. Let us now define $\phi_p(b, t) = 2b \sum_{n \in \text{odd}} \frac{\sin(\omega t_n)}{\omega} - 2b \sum_{n \in \text{even}} \frac{\sin(\omega t_n)}{\omega}$ with the $t_n < t$. Notice that this phase depends on the field amplitude $b$, the pulse sequence, and the time $t$. Then $e^{4ibC(t)} = e^{4i\phi_p(b,t)} e^{i4b(-1)^N \frac{\sin(\omega t)}{\omega}}$. We now choose the times $t_n$ such that $\omega t_n$ is an integer multiple of $\pi$ for all $t_i$. Interestingly, this is different from what is usually done using a single qubit with $b(t) = b \cos(\omega t)$ - in that case, the pulses are applied at times $t_i$ such that $\omega t_i  = (m + \frac{1}{2})\pi$ for integer $m$. For both $N$ even or odd, in the high frequency regime that we are dealing with, we can then write $e^{4iC(t)} \approx A$, where, as before, $A = J_0\left(\frac{4b}{\omega}\right)$. The effective Hamiltonian is then the same as Eq.~\eqref{a3}. As such, our scheme of estimating the oscillating field amplitude via the shift in the qubit-qubit interaction strength remains intact when pulses are applied to mitigate the effect of noise. Importantly, the time interval between the pulses need not be very small. This is illustrated in Fig.~\ref{labelnoiseandpulses}. For simplicity, here we have taken the Hamiltonian of the two qubits to be $H(t) = g\sigma^{(1)}_x \sigma^{(2)}_x +b \cos(\omega t) [\sigma^{(1)}_z +\sigma^{(2)}_z ] + B_x(t)[\sigma_x^{(1)} + \sigma_x^{(2)}] + B_y(t)[\sigma_y^{(1)} + \sigma_y^{(2)}] + B_z(t)[\sigma_z^{(1)} + \sigma_z^{(2)}]$, where the $B_m$ are independent random variables obtained by solving the Ornstein-Uhlenbeck equation cast in the form \cite{Jacobsbook} 
$$ dB_m = -\frac{(B_m - \mu)}{t_c} dt + \sigma \sqrt{\frac{2}{t_c}} dW. $$
Here $m = x, y, z$, $\mu$ is the mean, $\sigma$ is the standard deviation, $t_c$ is the correlation time, and $W$ is the standard Wiener process. Similar to the XY4 pulse sequence \cite{LidarPhysRevAppl2023}, we apply the $X$ pulses at times $n\delta t$ and the $Z$ pulses at times $(n + \frac{1}{2})\delta t$ for $n = 1, 2, 3, \hdots$. Importantly, $\delta t$ can be an integer multiple of $\frac{\pi}{\omega}$ (note that $\frac{\pi}{\omega}$ is the time interval between the pulses in the usual scheme where a pulse is applied whenever the field reverses direction). The dot-dashed blue curve in Fig.~\ref{labelnoiseandpulses} shows the dynamics of the expectation of the observable $M(t)$ in the presence of noise and the oscillating field, clearly illustrating the decay in $\langle M(t) \rangle$. However, in the presence of applied pulses (see the dashed, red curve), the dynamics are very close to those obtained in the presence of the oscillating field but in the absence of the noise (the solid, black curve). Therefore, as expected,  
the applied pulses greatly mitigate the effect of noise; in effect, the pulses increase the values of $T_1$ and $T_2$, thereby allowing us to approach the ideal results obtained in the absence of noise.

\section{Conclusion} \label{sec4}

In summary, we have introduced a scheme that utilizes two qubits (instead of one) to measure an oscillating field. The scheme relies on the change in the interaction strength between the two qubits due to the oscillating field. We showed this by first transforming to the frame of the oscillating field, and then deriving a time-independent effective Hamiltonian in this frame for the dynamics of the two qubits in the high-frequency regime. This effective Hamiltonian clearly shows that the interaction strength depends on the parameters of the oscillating field. We then showed how an observable could be measured to allow us to measure the oscillating field. In fact, we need to perform only single-qubit measurements. We computed the associated Fisher information to quantify the precision of the estimates. We showed that, even without any pulses, the Fisher information keeps increasing (in the absence of noise). Importantly, we computed the quantum Fisher information to demonstrate that we are performing the optimal measurement. We also showed that our scheme works even if we do not know the phase of the oscillating field. We then generalized our treatment to include noise and showed how pulses can suppress the Fisher information and thereby increase the precision of our estimates. To counter the effect of noise, we showed that pulses can still be applied, and our scheme will still work. However, crucially, the pulses need not be spaced very close to each other. Owing to great interest in quantum sensing in general and the measurement of oscillating fields in particular, our results should be beneficial in the metrology of high-frequency oscillating fields.  

\section*{Acknowledgements}
Support from the LUMS FIF grant FIF-0952 is acknowledged.

\section*{Data availability}

The computer programs and data that support the findings of this article are openly available \cite{data}.

\appendix 

\section{Modeling noise via a Lindblad master equation}
\label{appendixlindblad}

\begin{figure}[t!]
    \centering
    \includegraphics[width=0.45\textwidth]{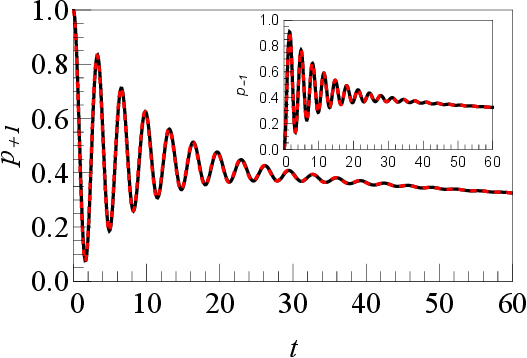}
    \caption{The main figure shows the behavior of the probability $p_{+1}$ as a function of time. The solid, black curve is $p_{+1}$ obtained by solving the Lindblad equation numerically. Here we are using dimensionless units with $\hbar = 1$ and $g = 1$, and we have $b = 1$, $\omega = 10$, $\gamma_1 = 0.01$ and $\gamma_2 = 0.05$. The dashed, red curve is obtained by fitting $p_{+1}$ to $\frac{1}{2} (1 + e^{-t/T_1})\frac{1}{2}\left[1 + e^{-t/T_2}\cos(2Agt)\right]$; we find $T_1 \approx 50$ and $T_2 \approx 10$. The figure in the inset is the same as the main figure, except that we are now plotting $p_{-1}$.}
    \label{labelnoisevialindblad}
\end{figure}

To model the effect of noise on our two-qubit sensor, we assume that each of the two qubits is subjected to amplitude damping as well as dephasing. As such, to model the dynamics of the sensor in the presence of the oscillating field, we use the Lindblad master equation given by
\[
\dot{\rho} \;=\; -\frac{i}{\hbar}\,[\widetilde{H},\rho] \;+\; \sum_{n} \left( C_n\,\rho\,C_n^\dagger \;-\; \frac{1}{2}\,\{C_n^\dagger C_n,\rho\} \right).
\]
Here $\widetilde{H}$ is the effective time-independent Hamiltonian of the two-qubit sensor, $C_1 = \sqrt{\gamma_1}\sigma_-^{(1)}$ and $C_2 = \sqrt{\gamma_1}\sigma_-^{(2)}$ lead to amplitude damping with the associated rate $\gamma_1$, and $C_3 = \sqrt{\gamma_2}\sigma_z^{(1)}$ and $C_4 = \sqrt{\gamma_2}\sigma_z^{(2)}$ lead to dephasing with the rate $\gamma_2$. The behavior of the probability of finding both the qubits in state $\ket{0}$ (that is, $p_{+1}$ in the notation followed in the main text) is illustrated in Fig.~\ref{labelnoisevialindblad}. With the chosen operators $C_1$, $C_2$, $C_3$, and $C_4$, we find that the state of our two-qubit sensor at long times becomes the maximally mixed state. We have fitted the probability $p_{+1}$ found numerically to a curve of the form $\frac{1}{2} (1 + e^{-t/T_1})\frac{1}{2}\left[1 + e^{-t/T_2}\cos(2Agt)\right]$. As should be obvious, the fit is excellent, thereby illustrating that we can indeed write down effectively that $p_{+1} =  \frac{1}{2} (1 + e^{-t/T_1})\frac{1}{2}\left[1 + e^{-t/T_2}\cos(2Agt)\right]$. Similarly, in the inset of Fig.~\ref{labelnoisevialindblad}, we have plotted the probability of finding both qubits in the state $\ket{1}$ (which is $p_{-1}$ in the notation followed in the main text), and fitted this to $\frac{1}{2} (1 + e^{-t/T_1})\frac{1}{2}\left[1 - e^{-t/T_2}\cos(2Agt)\right]$; once again, the fit is excellent. In short, we have justified the probabilities used in Sec.~\ref{sec3} to calculate the Fisher information for the estimation of the oscillating field amplitude in the presence of noise.

\section{Using single qubit measurements to find joint probabilities in the presence of noise}
\label{appendixmeasurement}

In the presence of noise, the state of our two-qubit sensor is not restricted to the subspace spanned by the states $\ket{00}$ and $\ket{11}$. However, we can use `nearly simultaneous' single qubit measurements to find $p_{+1}$ (a similar argument holds for $p_{-1}$). Let us suppose that the state of our two-qubit sensor at some arbitrary time is $\rho$. The probability that both qubits are in state $\ket{0}$ is given by $p_{+1} = \opav{00}{\rho}{00}$. As mentioned in the main text, such simultaneous measurements have been performed experimentally. Suppose, however, that we first measure qubit $1$ to be in the state $\ket{0}$, and then, after a short time interval $\delta t$, find qubit $2$ to be in the state $\ket{0}$ as well. We now show that the joint probability that both are in state $\ket{0}$, taking into account the time interval between the measurements, is approximately the same as $p_{+1}$, with the error being of order $(g\delta t)^2$. We start by noting that once we find qubit $1$ to be in the state $\ket{0}$, the state of the two qubits just after this measurement is proportional to $(\ket{0}\bra{0} \otimes \mathds{1})\rho(\ket{0}\bra{0}\otimes \mathds{1})$. We write this as $\alpha \ket{00}\bra{00} + \beta_1 \ket{00}\bra{01} + \beta_1^*\ket{01}\bra{00} + \beta_2 \ket{01}\bra{01}$, where $\alpha = \opav{00}{\rho}{00}$, $\beta_1 = \opav{00}{\rho}{01}$, and so on. Note that $\alpha$ is nothing other than $p_{+1}$. We normalize this state to obtain 
\begin{align*}
\rho^{(1)} = &\frac{1}{\alpha + \beta_2}(\alpha \ket{00}\bra{00} + \beta_1 \ket{00}\bra{01} + \\ 
&\beta_1^*\ket{01}\bra{00} + \beta_2 \ket{01}\bra{01}), 
\end{align*}
with the superscript in $\rho^{(1)}$ indicating that qubit 1 has been measured. Note that $(\alpha + \beta_2)$ is merely the probability that we found qubit $1$ in state $\ket{0}$. Now this state evolves over the time interval $\delta t$. We assume that this time interval is small so that over this time interval, the evolution of the state is essentially due to the effective time-independent Hamiltonian $\widetilde{H}$ - we ignore the effect of noise over this time interval. After all, for sensing, we need the effect of the noise to be much smaller than the effect of the coupling between the sensor and the quantity that we are trying to measure. The way forward is then clear, although laborious. We evolve the state $\rho^{(1)}$ with the Hamiltonian $\widetilde{H}$ over the time interval $\delta t$, and then measure qubit $2$. That is, we multiply the time-evolved state by $(\mathds{1}\otimes \ket{0}\bra{0})$, and then take the trace to find the probability that we measure qubit $2$ to be in the state $\ket{0}$, given that we found qubit $1$ to be in the state $\ket{0}$ as well. Carrying out this procedure, we find that this conditional probability is given by $\frac{\alpha \cos^2(gA \delta t) + \beta_2 \sin^2 (g \delta t)}{\alpha + \beta_2}$. The probability that we find both qubits to be in the state $\ket{0}$ is then given by $\alpha \cos^2(gA \delta t) + \beta_2 \sin^2 (g \delta t)$. If $\delta t$ is small such that $g \delta t \ll 1$, this joint probability is approximately $\alpha(1 - (g \delta t)^2 A^2) + (g \delta t)^2 \beta_2$. It is clear that this is approximately equal to $\alpha$ (which is $p_{+1}$); the error is of order $(g\delta t)^2$.


\begin{thebibliography}{39}%
\makeatletter
\providecommand \@ifxundefined [1]{%
 \@ifx{#1\undefined}
}%
\providecommand \@ifnum [1]{%
 \ifnum #1\expandafter \@firstoftwo
 \else \expandafter \@secondoftwo
 \fi
}%
\providecommand \@ifx [1]{%
 \ifx #1\expandafter \@firstoftwo
 \else \expandafter \@secondoftwo
 \fi
}%
\providecommand \natexlab [1]{#1}%
\providecommand \enquote  [1]{``#1''}%
\providecommand \bibnamefont  [1]{#1}%
\providecommand \bibfnamefont [1]{#1}%
\providecommand \citenamefont [1]{#1}%
\providecommand \href@noop [0]{\@secondoftwo}%
\providecommand \href [0]{\begingroup \@sanitize@url \@href}%
\providecommand \@href[1]{\@@startlink{#1}\@@href}%
\providecommand \@@href[1]{\endgroup#1\@@endlink}%
\providecommand \@sanitize@url [0]{\catcode `\\12\catcode `\$12\catcode
  `\&12\catcode `\#12\catcode `\^12\catcode `\_12\catcode `\%12\relax}%
\providecommand \@@startlink[1]{}%
\providecommand \@@endlink[0]{}%
\providecommand \url  [0]{\begingroup\@sanitize@url \@url }%
\providecommand \@url [1]{\endgroup\@href {#1}{\urlprefix }}%
\providecommand \urlprefix  [0]{URL }%
\providecommand \Eprint [0]{\href }%
\providecommand \doibase [0]{https://doi.org/}%
\providecommand \selectlanguage [0]{\@gobble}%
\providecommand \bibinfo  [0]{\@secondoftwo}%
\providecommand \bibfield  [0]{\@secondoftwo}%
\providecommand \translation [1]{[#1]}%
\providecommand \BibitemOpen [0]{}%
\providecommand \bibitemStop [0]{}%
\providecommand \bibitemNoStop [0]{.\EOS\space}%
\providecommand \EOS [0]{\spacefactor3000\relax}%
\providecommand \BibitemShut  [1]{\csname bibitem#1\endcsname}%
\let\auto@bib@innerbib\@empty
\bibitem [{\citenamefont {Aslam}\ \emph {et~al.}(2023)\citenamefont {Aslam},
  \citenamefont {Zhou}, \citenamefont {Urbach}, \citenamefont {Turner},
  \citenamefont {Walsworth}, \citenamefont {Lukin},\ and\ \citenamefont
  {Park}}]{AslamNatRevPhys2023}%
  \BibitemOpen
  \bibfield  {author} {\bibinfo {author} {\bibfnamefont {N.}~\bibnamefont
  {Aslam}}, \bibinfo {author} {\bibfnamefont {H.}~\bibnamefont {Zhou}},
  \bibinfo {author} {\bibfnamefont {E.~K.}\ \bibnamefont {Urbach}}, \bibinfo
  {author} {\bibfnamefont {M.~J.}\ \bibnamefont {Turner}}, \bibinfo {author}
  {\bibfnamefont {R.~L.}\ \bibnamefont {Walsworth}}, \bibinfo {author}
  {\bibfnamefont {M.~D.}\ \bibnamefont {Lukin}},\ and\ \bibinfo {author}
  {\bibfnamefont {H.}~\bibnamefont {Park}},\ }\bibfield  {title} {\bibinfo
  {title} {Quantum sensors for biomedical applications},\ }\href
  {https://doi.org/10.1038/s42254-023-00558-3} {\bibfield  {journal} {\bibinfo
  {journal} {Nat. Rev. Phys.}\ }\textbf {\bibinfo {volume} {5}},\ \bibinfo
  {pages} {157} (\bibinfo {year} {2023})}\BibitemShut {NoStop}%
\bibitem [{\citenamefont {Degen}\ \emph {et~al.}(2017)\citenamefont {Degen},
  \citenamefont {Reinhard},\ and\ \citenamefont
  {Cappellaro}}]{CappellaroRevModPhys2017}%
  \BibitemOpen
  \bibfield  {author} {\bibinfo {author} {\bibfnamefont {C.~L.}\ \bibnamefont
  {Degen}}, \bibinfo {author} {\bibfnamefont {F.}~\bibnamefont {Reinhard}},\
  and\ \bibinfo {author} {\bibfnamefont {P.}~\bibnamefont {Cappellaro}},\
  }\bibfield  {title} {\bibinfo {title} {Quantum sensing},\ }\href
  {https://doi.org/10.1103/RevModPhys.89.035002} {\bibfield  {journal}
  {\bibinfo  {journal} {Rev. Mod. Phys.}\ }\textbf {\bibinfo {volume} {89}},\
  \bibinfo {pages} {035002} (\bibinfo {year} {2017})}\BibitemShut {NoStop}%
\bibitem [{\citenamefont {Wang}\ \emph {et~al.}(2022)\citenamefont {Wang},
  \citenamefont {Liu}, \citenamefont {Schloss}, \citenamefont {Alsid},
  \citenamefont {Braje},\ and\ \citenamefont {Cappellaro}}]{CappellaroPRX2022}%
  \BibitemOpen
  \bibfield  {author} {\bibinfo {author} {\bibfnamefont {G.}~\bibnamefont
  {Wang}}, \bibinfo {author} {\bibfnamefont {Y.-X.}\ \bibnamefont {Liu}},
  \bibinfo {author} {\bibfnamefont {J.~M.}\ \bibnamefont {Schloss}}, \bibinfo
  {author} {\bibfnamefont {S.~T.}\ \bibnamefont {Alsid}}, \bibinfo {author}
  {\bibfnamefont {D.~A.}\ \bibnamefont {Braje}},\ and\ \bibinfo {author}
  {\bibfnamefont {P.}~\bibnamefont {Cappellaro}},\ }\bibfield  {title}
  {\bibinfo {title} {Sensing of arbitrary-frequency fields using a quantum
  mixer},\ }\href {https://doi.org/10.1103/PhysRevX.12.021061} {\bibfield
  {journal} {\bibinfo  {journal} {Phys. Rev. X}\ }\textbf {\bibinfo {volume}
  {12}},\ \bibinfo {pages} {021061} (\bibinfo {year} {2022})}\BibitemShut
  {NoStop}%
\bibitem [{\citenamefont {Bonizzoni}\ \emph {et~al.}(2024)\citenamefont
  {Bonizzoni}, \citenamefont {Ghirri}, \citenamefont {Santanni},\ and\
  \citenamefont {Affronte}}]{Bonizzonninpj2024}%
  \BibitemOpen
  \bibfield  {author} {\bibinfo {author} {\bibfnamefont {C.}~\bibnamefont
  {Bonizzoni}}, \bibinfo {author} {\bibfnamefont {A.}~\bibnamefont {Ghirri}},
  \bibinfo {author} {\bibfnamefont {F.}~\bibnamefont {Santanni}},\ and\
  \bibinfo {author} {\bibfnamefont {M.}~\bibnamefont {Affronte}},\ }\bibfield
  {title} {\bibinfo {title} {Quantum sensing of magnetic fields with molecular
  spins},\ }\href {https://doi.org/10.1038/s41534-024-00838-5} {\bibfield
  {journal} {\bibinfo  {journal} {npj Quantum Inf}\ }\textbf {\bibinfo {volume}
  {10}},\ \bibinfo {pages} {41} (\bibinfo {year} {2024})}\BibitemShut {NoStop}%
\bibitem [{\citenamefont {Iemini}\ \emph {et~al.}(2024)\citenamefont {Iemini},
  \citenamefont {Fazio},\ and\ \citenamefont {Sanpera}}]{SanperaPRA2024}%
  \BibitemOpen
  \bibfield  {author} {\bibinfo {author} {\bibfnamefont {F.}~\bibnamefont
  {Iemini}}, \bibinfo {author} {\bibfnamefont {R.}~\bibnamefont {Fazio}},\ and\
  \bibinfo {author} {\bibfnamefont {A.}~\bibnamefont {Sanpera}},\ }\bibfield
  {title} {\bibinfo {title} {Floquet time crystals as quantum sensors of ac
  fields},\ }\href {https://doi.org/10.1103/PhysRevA.109.L050203} {\bibfield
  {journal} {\bibinfo  {journal} {Phys. Rev. A}\ }\textbf {\bibinfo {volume}
  {109}},\ \bibinfo {pages} {L050203} (\bibinfo {year} {2024})}\BibitemShut
  {NoStop}%
\bibitem [{\citenamefont {Freeman}\ and\ \citenamefont
  {Choi}(2001)}]{FreemanScience2001}%
  \BibitemOpen
  \bibfield  {author} {\bibinfo {author} {\bibfnamefont {M.~R.}\ \bibnamefont
  {Freeman}}\ and\ \bibinfo {author} {\bibfnamefont {B.~C.}\ \bibnamefont
  {Choi}},\ }\href@noop {} {\bibfield  {journal} {\bibinfo  {journal}
  {Science}\ }\textbf {\bibinfo {volume} {294}},\ \bibinfo {pages} {1484}
  (\bibinfo {year} {2001})}\BibitemShut {NoStop}%
\bibitem [{\citenamefont {Taylor}\ \emph {et~al.}(2008)\citenamefont {Taylor},
  \citenamefont {Cappellaro}, \citenamefont {Childress}, \citenamefont {Jiang},
  \citenamefont {Budker}, \citenamefont {Hemmer}, \citenamefont {Yacoby},
  \citenamefont {Walsworth},\ and\ \citenamefont {Lukin}}]{TaylorNatPhys2008}%
  \BibitemOpen
  \bibfield  {author} {\bibinfo {author} {\bibfnamefont {J.~M.}\ \bibnamefont
  {Taylor}}, \bibinfo {author} {\bibfnamefont {P.}~\bibnamefont {Cappellaro}},
  \bibinfo {author} {\bibfnamefont {L.}~\bibnamefont {Childress}}, \bibinfo
  {author} {\bibfnamefont {L.}~\bibnamefont {Jiang}}, \bibinfo {author}
  {\bibfnamefont {D.}~\bibnamefont {Budker}}, \bibinfo {author} {\bibfnamefont
  {P.~R.}\ \bibnamefont {Hemmer}}, \bibinfo {author} {\bibfnamefont
  {A.}~\bibnamefont {Yacoby}}, \bibinfo {author} {\bibfnamefont
  {R.}~\bibnamefont {Walsworth}},\ and\ \bibinfo {author} {\bibfnamefont
  {M.~D.}\ \bibnamefont {Lukin}},\ }\href@noop {} {\bibfield  {journal}
  {\bibinfo  {journal} {Nat.~Phys.}\ }\textbf {\bibinfo {volume} {4}},\
  \bibinfo {pages} {810} (\bibinfo {year} {2008})}\BibitemShut {NoStop}%
\bibitem [{\citenamefont {Maze}\ \emph {et~al.}(2008)\citenamefont {Maze},
  \citenamefont {Stanwix}, \citenamefont {Hodges}, \citenamefont {Hong},
  \citenamefont {Taylor}, \citenamefont {Cappellaro}, \citenamefont {Jiang},
  \citenamefont {Dutt}, \citenamefont {Togan}, \citenamefont {Zibrov},
  \citenamefont {Yacoby}, \citenamefont {Walsworth},\ and\ \citenamefont
  {Lukin}}]{MazeNature2008}%
  \BibitemOpen
  \bibfield  {author} {\bibinfo {author} {\bibfnamefont {J.~R.}\ \bibnamefont
  {Maze}}, \bibinfo {author} {\bibfnamefont {P.~L.}\ \bibnamefont {Stanwix}},
  \bibinfo {author} {\bibfnamefont {J.~S.}\ \bibnamefont {Hodges}}, \bibinfo
  {author} {\bibfnamefont {S.}~\bibnamefont {Hong}}, \bibinfo {author}
  {\bibfnamefont {J.~M.}\ \bibnamefont {Taylor}}, \bibinfo {author}
  {\bibfnamefont {P.}~\bibnamefont {Cappellaro}}, \bibinfo {author}
  {\bibfnamefont {L.}~\bibnamefont {Jiang}}, \bibinfo {author} {\bibfnamefont
  {M.~V.~G.}\ \bibnamefont {Dutt}}, \bibinfo {author} {\bibfnamefont
  {E.}~\bibnamefont {Togan}}, \bibinfo {author} {\bibfnamefont {A.~S.}\
  \bibnamefont {Zibrov}}, \bibinfo {author} {\bibfnamefont {A.}~\bibnamefont
  {Yacoby}}, \bibinfo {author} {\bibfnamefont {R.~L.}\ \bibnamefont
  {Walsworth}},\ and\ \bibinfo {author} {\bibfnamefont {M.~D.}\ \bibnamefont
  {Lukin}},\ }\href@noop {} {\bibfield  {journal} {\bibinfo  {journal}
  {Nature}\ }\textbf {\bibinfo {volume} {455}},\ \bibinfo {pages} {644}
  (\bibinfo {year} {2008})}\BibitemShut {NoStop}%
\bibitem [{\citenamefont {Balasubramanian}\ \emph {et~al.}(2008)\citenamefont
  {Balasubramanian}, \citenamefont {Chan}, \citenamefont {Kolesov},
  \citenamefont {Al-Hmoud}, \citenamefont {Tisler}, \citenamefont {Shin},
  \citenamefont {Kim}, \citenamefont {Wojcik}, \citenamefont {Hemmer},
  \citenamefont {Krueger}, \citenamefont {Hanke}, \citenamefont
  {Leitenstorfer}, \citenamefont {Bratschitsch}, \citenamefont {Jelezko},\ and\
  \citenamefont {Wrachtrup}}]{BalasubramanianNature2008}%
  \BibitemOpen
  \bibfield  {author} {\bibinfo {author} {\bibfnamefont {G.}~\bibnamefont
  {Balasubramanian}}, \bibinfo {author} {\bibfnamefont {I.~Y.}\ \bibnamefont
  {Chan}}, \bibinfo {author} {\bibfnamefont {R.}~\bibnamefont {Kolesov}},
  \bibinfo {author} {\bibfnamefont {M.}~\bibnamefont {Al-Hmoud}}, \bibinfo
  {author} {\bibfnamefont {J.}~\bibnamefont {Tisler}}, \bibinfo {author}
  {\bibfnamefont {C.}~\bibnamefont {Shin}}, \bibinfo {author} {\bibfnamefont
  {C.}~\bibnamefont {Kim}}, \bibinfo {author} {\bibfnamefont {A.}~\bibnamefont
  {Wojcik}}, \bibinfo {author} {\bibfnamefont {P.~R.}\ \bibnamefont {Hemmer}},
  \bibinfo {author} {\bibfnamefont {A.}~\bibnamefont {Krueger}}, \bibinfo
  {author} {\bibfnamefont {T.}~\bibnamefont {Hanke}}, \bibinfo {author}
  {\bibfnamefont {A.}~\bibnamefont {Leitenstorfer}}, \bibinfo {author}
  {\bibfnamefont {R.}~\bibnamefont {Bratschitsch}}, \bibinfo {author}
  {\bibfnamefont {F.}~\bibnamefont {Jelezko}},\ and\ \bibinfo {author}
  {\bibfnamefont {J.}~\bibnamefont {Wrachtrup}},\ }\href@noop {} {\bibfield
  {journal} {\bibinfo  {journal} {Nature}\ }\textbf {\bibinfo {volume} {455}},\
  \bibinfo {pages} {648} (\bibinfo {year} {2008})}\BibitemShut {NoStop}%
\bibitem [{\citenamefont {Chang}\ \emph {et~al.}(2008)\citenamefont {Chang},
  \citenamefont {Lee}, \citenamefont {Chen}, \citenamefont {Chang},
  \citenamefont {Tsai}, \citenamefont {Fu}, \citenamefont {Lim}, \citenamefont
  {Tzeng}, \citenamefont {Fang}, \citenamefont {Han}, \citenamefont {Chang},\
  and\ \citenamefont {Fann}}]{ChangNatNano2008}%
  \BibitemOpen
  \bibfield  {author} {\bibinfo {author} {\bibfnamefont {Y.-R.}\ \bibnamefont
  {Chang}}, \bibinfo {author} {\bibfnamefont {H.-Y.}\ \bibnamefont {Lee}},
  \bibinfo {author} {\bibfnamefont {K.}~\bibnamefont {Chen}}, \bibinfo {author}
  {\bibfnamefont {C.-C.}\ \bibnamefont {Chang}}, \bibinfo {author}
  {\bibfnamefont {D.-S.}\ \bibnamefont {Tsai}}, \bibinfo {author}
  {\bibfnamefont {C.-C.}\ \bibnamefont {Fu}}, \bibinfo {author} {\bibfnamefont
  {T.-S.}\ \bibnamefont {Lim}}, \bibinfo {author} {\bibfnamefont {Y.-K.}\
  \bibnamefont {Tzeng}}, \bibinfo {author} {\bibfnamefont {C.-Y.}\ \bibnamefont
  {Fang}}, \bibinfo {author} {\bibfnamefont {C.-C.}\ \bibnamefont {Han}},
  \bibinfo {author} {\bibfnamefont {H.-C.}\ \bibnamefont {Chang}},\ and\
  \bibinfo {author} {\bibfnamefont {W.}~\bibnamefont {Fann}},\ }\bibfield
  {title} {\bibinfo {title} {Mass production and dynamic imaging of fluorescent
  nanodiamonds},\ }\href@noop {} {\bibfield  {journal} {\bibinfo  {journal}
  {Nat.~Nanotechnol.}\ }\textbf {\bibinfo {volume} {3}},\ \bibinfo {pages}
  {284} (\bibinfo {year} {2008})}\BibitemShut {NoStop}%
\bibitem [{\citenamefont {Balasubramanian}\ \emph {et~al.}(2009)\citenamefont
  {Balasubramanian}, \citenamefont {Neumann}, \citenamefont {Twitchen},
  \citenamefont {Markham}, \citenamefont {Kolesov}, \citenamefont {Mizuochi},
  \citenamefont {Isoya}, \citenamefont {Achard}, \citenamefont {Beck},
  \citenamefont {Tissler}, \citenamefont {Jacques}, \citenamefont {Hemmer},
  \citenamefont {Jelezko},\ and\ \citenamefont
  {Wrachtrup}}]{BalasubramanianNatNano2009}%
  \BibitemOpen
  \bibfield  {author} {\bibinfo {author} {\bibfnamefont {G.}~\bibnamefont
  {Balasubramanian}}, \bibinfo {author} {\bibfnamefont {P.}~\bibnamefont
  {Neumann}}, \bibinfo {author} {\bibfnamefont {D.}~\bibnamefont {Twitchen}},
  \bibinfo {author} {\bibfnamefont {M.}~\bibnamefont {Markham}}, \bibinfo
  {author} {\bibfnamefont {R.}~\bibnamefont {Kolesov}}, \bibinfo {author}
  {\bibfnamefont {N.}~\bibnamefont {Mizuochi}}, \bibinfo {author}
  {\bibfnamefont {J.}~\bibnamefont {Isoya}}, \bibinfo {author} {\bibfnamefont
  {J.}~\bibnamefont {Achard}}, \bibinfo {author} {\bibfnamefont
  {J.}~\bibnamefont {Beck}}, \bibinfo {author} {\bibfnamefont {J.}~\bibnamefont
  {Tissler}}, \bibinfo {author} {\bibfnamefont {V.}~\bibnamefont {Jacques}},
  \bibinfo {author} {\bibfnamefont {P.~R.}\ \bibnamefont {Hemmer}}, \bibinfo
  {author} {\bibfnamefont {F.}~\bibnamefont {Jelezko}},\ and\ \bibinfo {author}
  {\bibfnamefont {J.}~\bibnamefont {Wrachtrup}},\ }\bibfield  {title} {\bibinfo
  {title} {Ultralong spin coherence time in isotopically engineered diamond},\
  }\href@noop {} {\bibfield  {journal} {\bibinfo  {journal} {Nat. Mater.}\
  }\textbf {\bibinfo {volume} {8}},\ \bibinfo {pages} {383} (\bibinfo {year}
  {2009})}\BibitemShut {NoStop}%
\bibitem [{\citenamefont {McGuinness}\ \emph {et~al.}(2011)\citenamefont
  {McGuinness}, \citenamefont {Yan}, \citenamefont {Stacey}, \citenamefont
  {Simpson}, \citenamefont {Hall}, \citenamefont {Maclaurin}, \citenamefont
  {Prawer}, \citenamefont {Mulvaney}, \citenamefont {Wrachtrup}, \citenamefont
  {Caruso}, \citenamefont {Scholten},\ and\ \citenamefont
  {Hollenberg}}]{McguinnessNatNano2011}%
  \BibitemOpen
  \bibfield  {author} {\bibinfo {author} {\bibfnamefont {L.~P.}\ \bibnamefont
  {McGuinness}}, \bibinfo {author} {\bibfnamefont {Y.}~\bibnamefont {Yan}},
  \bibinfo {author} {\bibfnamefont {A.}~\bibnamefont {Stacey}}, \bibinfo
  {author} {\bibfnamefont {D.~A.}\ \bibnamefont {Simpson}}, \bibinfo {author}
  {\bibfnamefont {L.~T.}\ \bibnamefont {Hall}}, \bibinfo {author}
  {\bibfnamefont {D.}~\bibnamefont {Maclaurin}}, \bibinfo {author}
  {\bibfnamefont {S.}~\bibnamefont {Prawer}}, \bibinfo {author} {\bibfnamefont
  {P.}~\bibnamefont {Mulvaney}}, \bibinfo {author} {\bibfnamefont
  {J.}~\bibnamefont {Wrachtrup}}, \bibinfo {author} {\bibfnamefont
  {F.}~\bibnamefont {Caruso}}, \bibinfo {author} {\bibfnamefont {R.~E.}\
  \bibnamefont {Scholten}},\ and\ \bibinfo {author} {\bibfnamefont {L.~C.~L.}\
  \bibnamefont {Hollenberg}},\ }\bibfield  {title} {\bibinfo {title} {Quantum
  measurement and orientation tracking of fluorescent nanodiamonds inside
  living cells},\ }\href@noop {} {\bibfield  {journal} {\bibinfo  {journal}
  {Nat.~Nanotechnol.}\ }\textbf {\bibinfo {volume} {6}},\ \bibinfo {pages}
  {358} (\bibinfo {year} {2011})}\BibitemShut {NoStop}%
\bibitem [{\citenamefont {de~Lange}\ \emph {et~al.}(2011)\citenamefont
  {de~Lange}, \citenamefont {Rist\`e}, \citenamefont {Dobrovitski},\ and\
  \citenamefont {Hanson}}]{deLangePRL2011}%
  \BibitemOpen
  \bibfield  {author} {\bibinfo {author} {\bibfnamefont {G.}~\bibnamefont
  {de~Lange}}, \bibinfo {author} {\bibfnamefont {D.}~\bibnamefont {Rist\`e}},
  \bibinfo {author} {\bibfnamefont {V.~V.}\ \bibnamefont {Dobrovitski}},\ and\
  \bibinfo {author} {\bibfnamefont {R.}~\bibnamefont {Hanson}},\ }\bibfield
  {title} {\bibinfo {title} {Single-spin magnetometry with multipulse sensing
  sequences},\ }\href {https://doi.org/10.1103/PhysRevLett.106.080802}
  {\bibfield  {journal} {\bibinfo  {journal} {Phys. Rev. Lett.}\ }\textbf
  {\bibinfo {volume} {106}},\ \bibinfo {pages} {080802} (\bibinfo {year}
  {2011})}\BibitemShut {NoStop}%
\bibitem [{\citenamefont {Horowitz}\ \emph {et~al.}(2012)\citenamefont
  {Horowitz}, \citenamefont {Alem{\'a}n}, \citenamefont {Christle},
  \citenamefont {Cleland},\ and\ \citenamefont {Awschalom}}]{HorowitzPNAS2012}%
  \BibitemOpen
  \bibfield  {author} {\bibinfo {author} {\bibfnamefont {V.~R.}\ \bibnamefont
  {Horowitz}}, \bibinfo {author} {\bibfnamefont {B.~J.}\ \bibnamefont
  {Alem{\'a}n}}, \bibinfo {author} {\bibfnamefont {D.~J.}\ \bibnamefont
  {Christle}}, \bibinfo {author} {\bibfnamefont {A.~N.}\ \bibnamefont
  {Cleland}},\ and\ \bibinfo {author} {\bibfnamefont {D.~D.}\ \bibnamefont
  {Awschalom}},\ }\bibfield  {title} {\bibinfo {title} {Electron spin resonance
  of nitrogen-vacancy centers in optically trapped nanodiamonds},\ }\href@noop
  {} {\bibfield  {journal} {\bibinfo  {journal} {Proc. Nat. Acad. Sci.}\
  }\textbf {\bibinfo {volume} {109}},\ \bibinfo {pages} {13493} (\bibinfo
  {year} {2012})}\BibitemShut {NoStop}%
\bibitem [{\citenamefont {Hirose}\ \emph {et~al.}(2012)\citenamefont {Hirose},
  \citenamefont {Aiello},\ and\ \citenamefont {Cappellaro}}]{HirosePRA2012}%
  \BibitemOpen
  \bibfield  {author} {\bibinfo {author} {\bibfnamefont {M.}~\bibnamefont
  {Hirose}}, \bibinfo {author} {\bibfnamefont {C.~D.}\ \bibnamefont {Aiello}},\
  and\ \bibinfo {author} {\bibfnamefont {P.}~\bibnamefont {Cappellaro}},\
  }\bibfield  {title} {\bibinfo {title} {Continuous dynamical decoupling
  magnetometry},\ }\href {https://doi.org/10.1103/PhysRevA.86.062320}
  {\bibfield  {journal} {\bibinfo  {journal} {Phys. Rev. A}\ }\textbf {\bibinfo
  {volume} {86}},\ \bibinfo {pages} {062320} (\bibinfo {year}
  {2012})}\BibitemShut {NoStop}%
\bibitem [{\citenamefont {Hall}\ \emph {et~al.}(2013)\citenamefont {Hall},
  \citenamefont {Simpson},\ and\ \citenamefont {Hollenberg}}]{HallMRS2013}%
  \BibitemOpen
  \bibfield  {author} {\bibinfo {author} {\bibfnamefont {L.}~\bibnamefont
  {Hall}}, \bibinfo {author} {\bibfnamefont {D.}~\bibnamefont {Simpson}},\ and\
  \bibinfo {author} {\bibfnamefont {L.}~\bibnamefont {Hollenberg}},\ }\bibfield
   {title} {\bibinfo {title} {Nanoscale sensing and imaging in biology using
  the nitrogen-vacancy center in diamond},\ }\href
  {https://doi.org/10.1557/mrs.2013.24} {\bibfield  {journal} {\bibinfo
  {journal} {MRS Bull.}\ }\textbf {\bibinfo {volume} {38}},\ \bibinfo {pages}
  {162} (\bibinfo {year} {2013})}\BibitemShut {NoStop}%
\bibitem [{\citenamefont {Hong}\ \emph {et~al.}(2013)\citenamefont {Hong},
  \citenamefont {Grinolds}, \citenamefont {Pham}, \citenamefont {Le~Sage},
  \citenamefont {Luan}, \citenamefont {Walsworth},\ and\ \citenamefont
  {Yacoby}}]{HongMRS2013}%
  \BibitemOpen
  \bibfield  {author} {\bibinfo {author} {\bibfnamefont {S.}~\bibnamefont
  {Hong}}, \bibinfo {author} {\bibfnamefont {M.~S.}\ \bibnamefont {Grinolds}},
  \bibinfo {author} {\bibfnamefont {L.~M.}\ \bibnamefont {Pham}}, \bibinfo
  {author} {\bibfnamefont {D.}~\bibnamefont {Le~Sage}}, \bibinfo {author}
  {\bibfnamefont {L.}~\bibnamefont {Luan}}, \bibinfo {author} {\bibfnamefont
  {R.~L.}\ \bibnamefont {Walsworth}},\ and\ \bibinfo {author} {\bibfnamefont
  {A.}~\bibnamefont {Yacoby}},\ }\bibfield  {title} {\bibinfo {title}
  {Nanoscale magnetometry with nv centers in diamond},\ }\href
  {https://doi.org/10.1557/mrs.2013.23} {\bibfield  {journal} {\bibinfo
  {journal} {MRS Bull.}\ }\textbf {\bibinfo {volume} {38}},\ \bibinfo {pages}
  {155} (\bibinfo {year} {2013})}\BibitemShut {NoStop}%
\bibitem [{\citenamefont {Nusran}\ and\ \citenamefont
  {Dutt}(2013)}]{NusranPRB2013}%
  \BibitemOpen
  \bibfield  {author} {\bibinfo {author} {\bibfnamefont {N.~M.}\ \bibnamefont
  {Nusran}}\ and\ \bibinfo {author} {\bibfnamefont {M.~V.~G.}\ \bibnamefont
  {Dutt}},\ }\bibfield  {title} {\bibinfo {title} {Dual-channel lock-in
  magnetometer with a single spin in diamond},\ }\href
  {https://doi.org/10.1103/PhysRevB.88.220410} {\bibfield  {journal} {\bibinfo
  {journal} {Phys. Rev. B}\ }\textbf {\bibinfo {volume} {88}},\ \bibinfo
  {pages} {220410} (\bibinfo {year} {2013})}\BibitemShut {NoStop}%
\bibitem [{\citenamefont {Loretz}\ \emph {et~al.}(2013)\citenamefont {Loretz},
  \citenamefont {Rosskopf},\ and\ \citenamefont {Degen}}]{LoretzPRL2013}%
  \BibitemOpen
  \bibfield  {author} {\bibinfo {author} {\bibfnamefont {M.}~\bibnamefont
  {Loretz}}, \bibinfo {author} {\bibfnamefont {T.}~\bibnamefont {Rosskopf}},\
  and\ \bibinfo {author} {\bibfnamefont {C.~L.}\ \bibnamefont {Degen}},\
  }\bibfield  {title} {\bibinfo {title} {Radio-frequency magnetometry using a
  single electron spin},\ }\href
  {https://doi.org/10.1103/PhysRevLett.110.017602} {\bibfield  {journal}
  {\bibinfo  {journal} {Phys. Rev. Lett.}\ }\textbf {\bibinfo {volume} {110}},\
  \bibinfo {pages} {017602} (\bibinfo {year} {2013})}\BibitemShut {NoStop}%
\bibitem [{\citenamefont {Le~Sage}\ \emph {et~al.}(2013)\citenamefont
  {Le~Sage}, \citenamefont {Arai}, \citenamefont {Glenn}, \citenamefont
  {DeVience}, \citenamefont {Pham}, \citenamefont {Rahn-Lee}, \citenamefont
  {Lukin}, \citenamefont {Yacoby}, \citenamefont {Komeili},\ and\ \citenamefont
  {Walsworth}}]{LeSageNature2013}%
  \BibitemOpen
  \bibfield  {author} {\bibinfo {author} {\bibfnamefont {D.}~\bibnamefont
  {Le~Sage}}, \bibinfo {author} {\bibfnamefont {K.}~\bibnamefont {Arai}},
  \bibinfo {author} {\bibfnamefont {D.~R.}\ \bibnamefont {Glenn}}, \bibinfo
  {author} {\bibfnamefont {S.~J.}\ \bibnamefont {DeVience}}, \bibinfo {author}
  {\bibfnamefont {L.~M.}\ \bibnamefont {Pham}}, \bibinfo {author}
  {\bibfnamefont {L.}~\bibnamefont {Rahn-Lee}}, \bibinfo {author}
  {\bibfnamefont {M.~D.}\ \bibnamefont {Lukin}}, \bibinfo {author}
  {\bibfnamefont {A.}~\bibnamefont {Yacoby}}, \bibinfo {author} {\bibfnamefont
  {A.}~\bibnamefont {Komeili}},\ and\ \bibinfo {author} {\bibfnamefont {R.~L.}\
  \bibnamefont {Walsworth}},\ }\bibfield  {title} {\bibinfo {title} {Optical
  magnetic imaging of living cells},\ }\href@noop {} {\bibfield  {journal}
  {\bibinfo  {journal} {Nature}\ }\textbf {\bibinfo {volume} {496}},\ \bibinfo
  {pages} {486} (\bibinfo {year} {2013})}\BibitemShut {NoStop}%
\bibitem [{\citenamefont {Geiselmann}\ \emph {et~al.}(2013)\citenamefont
  {Geiselmann}, \citenamefont {Juan}, \citenamefont {Renger}, \citenamefont
  {Say}, \citenamefont {Brown}, \citenamefont {de~Abajo}, \citenamefont
  {Koppens},\ and\ \citenamefont {Quidant}}]{GeiselmannNatureNano2013}%
  \BibitemOpen
  \bibfield  {author} {\bibinfo {author} {\bibfnamefont {M.}~\bibnamefont
  {Geiselmann}}, \bibinfo {author} {\bibfnamefont {M.~L.}\ \bibnamefont
  {Juan}}, \bibinfo {author} {\bibfnamefont {J.}~\bibnamefont {Renger}},
  \bibinfo {author} {\bibfnamefont {J.~M.}\ \bibnamefont {Say}}, \bibinfo
  {author} {\bibfnamefont {L.~J.}\ \bibnamefont {Brown}}, \bibinfo {author}
  {\bibfnamefont {F.~J.~G.}\ \bibnamefont {de~Abajo}}, \bibinfo {author}
  {\bibfnamefont {F.}~\bibnamefont {Koppens}},\ and\ \bibinfo {author}
  {\bibfnamefont {R.}~\bibnamefont {Quidant}},\ }\bibfield  {title} {\bibinfo
  {title} {Three-dimensional optical manipulation of a single electron spin},\
  }\href@noop {} {\bibfield  {journal} {\bibinfo  {journal} {Nat.
  Nanotechnol.}\ }\textbf {\bibinfo {volume} {8}},\ \bibinfo {pages} {175}
  (\bibinfo {year} {2013})}\BibitemShut {NoStop}%
\bibitem [{\citenamefont {Magesan}\ \emph {et~al.}(2013)\citenamefont
  {Magesan}, \citenamefont {Cooper}, \citenamefont {Yum},\ and\ \citenamefont
  {Cappellaro}}]{MagesanPRA2013}%
  \BibitemOpen
  \bibfield  {author} {\bibinfo {author} {\bibfnamefont {E.}~\bibnamefont
  {Magesan}}, \bibinfo {author} {\bibfnamefont {A.}~\bibnamefont {Cooper}},
  \bibinfo {author} {\bibfnamefont {H.}~\bibnamefont {Yum}},\ and\ \bibinfo
  {author} {\bibfnamefont {P.}~\bibnamefont {Cappellaro}},\ }\bibfield  {title}
  {\bibinfo {title} {Reconstructing the profile of time-varying magnetic fields
  with quantum sensors},\ }\href {https://doi.org/10.1103/PhysRevA.88.032107}
  {\bibfield  {journal} {\bibinfo  {journal} {Phys. Rev. A}\ }\textbf {\bibinfo
  {volume} {88}},\ \bibinfo {pages} {032107} (\bibinfo {year}
  {2013})}\BibitemShut {NoStop}%
\bibitem [{\citenamefont {Cooper}\ \emph {et~al.}(2014)\citenamefont {Cooper},
  \citenamefont {Magesan}, \citenamefont {Yum},\ and\ \citenamefont
  {Cappellaro}}]{CooperNatCommun2014}%
  \BibitemOpen
  \bibfield  {author} {\bibinfo {author} {\bibfnamefont {A.}~\bibnamefont
  {Cooper}}, \bibinfo {author} {\bibfnamefont {E.}~\bibnamefont {Magesan}},
  \bibinfo {author} {\bibfnamefont {H.~N.}\ \bibnamefont {Yum}},\ and\ \bibinfo
  {author} {\bibfnamefont {P.}~\bibnamefont {Cappellaro}},\ }\href@noop {}
  {\bibfield  {journal} {\bibinfo  {journal} {Nat. Commun.}\ }\textbf {\bibinfo
  {volume} {5}},\ \bibinfo {pages} {3141} (\bibinfo {year} {2014})}\BibitemShut
  {NoStop}%
\bibitem [{\citenamefont {Nusran}\ and\ \citenamefont
  {Dutt}(2014)}]{NusranarXiv2014}%
  \BibitemOpen
  \bibfield  {author} {\bibinfo {author} {\bibfnamefont {N.~M.}\ \bibnamefont
  {Nusran}}\ and\ \bibinfo {author} {\bibfnamefont {M.~V.~G.}\ \bibnamefont
  {Dutt}},\ }\href@noop {} {\bibfield  {journal} {\bibinfo  {journal} {e-print
  arXiv:1403.4506}\ } (\bibinfo {year} {2014})}\BibitemShut {NoStop}%
\bibitem [{\citenamefont {Schirhagl}\ \emph {et~al.}(2014)\citenamefont
  {Schirhagl}, \citenamefont {Chang}, \citenamefont {Loretz},\ and\
  \citenamefont {Degen}}]{DegenRev2014}%
  \BibitemOpen
  \bibfield  {author} {\bibinfo {author} {\bibfnamefont {R.}~\bibnamefont
  {Schirhagl}}, \bibinfo {author} {\bibfnamefont {K.}~\bibnamefont {Chang}},
  \bibinfo {author} {\bibfnamefont {M.}~\bibnamefont {Loretz}},\ and\ \bibinfo
  {author} {\bibfnamefont {C.~L.}\ \bibnamefont {Degen}},\ }\bibfield  {title}
  {\bibinfo {title} {Nitrogen-vacancy centers in diamond: Nanoscale sensors for
  physics and biology},\ }\href
  {https://doi.org/https://doi.org/10.1146/annurev-physchem-040513-103659}
  {\bibfield  {journal} {\bibinfo  {journal} {Annual Review of Physical
  Chemistry}\ }\textbf {\bibinfo {volume} {65}},\ \bibinfo {pages} {83}
  (\bibinfo {year} {2014})}\BibitemShut {NoStop}%
\bibitem [{\citenamefont {Rondin}\ \emph {et~al.}(2014)\citenamefont {Rondin},
  \citenamefont {Tetienne}, \citenamefont {Hingant}, \citenamefont {Roch},
  \citenamefont {Maletinsky},\ and\ \citenamefont {Jacques}}]{RondinRev2014}%
  \BibitemOpen
  \bibfield  {author} {\bibinfo {author} {\bibfnamefont {L.}~\bibnamefont
  {Rondin}}, \bibinfo {author} {\bibfnamefont {J.-P.}\ \bibnamefont
  {Tetienne}}, \bibinfo {author} {\bibfnamefont {T.}~\bibnamefont {Hingant}},
  \bibinfo {author} {\bibfnamefont {J.-F.}\ \bibnamefont {Roch}}, \bibinfo
  {author} {\bibfnamefont {P.}~\bibnamefont {Maletinsky}},\ and\ \bibinfo
  {author} {\bibfnamefont {V.}~\bibnamefont {Jacques}},\ }\bibfield  {title}
  {\bibinfo {title} {Magnetometry with nitrogen-vacancy defects in diamond},\
  }\href {https://doi.org/10.1088/0034-4885/77/5/056503} {\bibfield  {journal}
  {\bibinfo  {journal} {Reports on Progress in Physics}\ }\textbf {\bibinfo
  {volume} {77}},\ \bibinfo {pages} {056503} (\bibinfo {year}
  {2014})}\BibitemShut {NoStop}%
\bibitem [{\citenamefont {Chaudhry}(2014)}]{c28}%
  \BibitemOpen
  \bibfield  {author} {\bibinfo {author} {\bibfnamefont {A.~Z.}\ \bibnamefont
  {Chaudhry}},\ }\bibfield  {title} {\bibinfo {title} {Utilizing
  nitrogen-vacancy centers to measure oscillating magnetic fields},\ }\href
  {https://doi.org/10.1103/PhysRevA.90.042104} {\bibfield  {journal} {\bibinfo
  {journal} {Phys. Rev. A}\ }\textbf {\bibinfo {volume} {90}},\ \bibinfo
  {pages} {042104} (\bibinfo {year} {2014})}\BibitemShut {NoStop}%
\bibitem [{\citenamefont {Chaudhry}(2015)}]{ChaudhryPRA2015}%
  \BibitemOpen
  \bibfield  {author} {\bibinfo {author} {\bibfnamefont {A.~Z.}\ \bibnamefont
  {Chaudhry}},\ }\bibfield  {title} {\bibinfo {title} {Detecting the presence
  of weak magnetic fields using nitrogen-vacancy centers},\ }\href
  {https://doi.org/10.1103/PhysRevA.91.062111} {\bibfield  {journal} {\bibinfo
  {journal} {Phys. Rev. A}\ }\textbf {\bibinfo {volume} {91}},\ \bibinfo
  {pages} {062111} (\bibinfo {year} {2015})}\BibitemShut {NoStop}%
\bibitem [{\citenamefont {Suter}\ and\ \citenamefont
  {\'Alvarez}(2016)}]{SuterRev2016}%
  \BibitemOpen
  \bibfield  {author} {\bibinfo {author} {\bibfnamefont {D.}~\bibnamefont
  {Suter}}\ and\ \bibinfo {author} {\bibfnamefont {G.~A.}\ \bibnamefont
  {\'Alvarez}},\ }\bibfield  {title} {\bibinfo {title} {Colloquium: Protecting
  quantum information against environmental noise},\ }\href
  {https://doi.org/10.1103/RevModPhys.88.041001} {\bibfield  {journal}
  {\bibinfo  {journal} {Rev. Mod. Phys.}\ }\textbf {\bibinfo {volume} {88}},\
  \bibinfo {pages} {041001} (\bibinfo {year} {2016})}\BibitemShut {NoStop}%
\bibitem [{\citenamefont {Pham}\ \emph {et~al.}(2012)\citenamefont {Pham},
  \citenamefont {Bar-Gill}, \citenamefont {Belthangady}, \citenamefont
  {Le~Sage}, \citenamefont {Cappellaro}, \citenamefont {Lukin}, \citenamefont
  {Yacoby},\ and\ \citenamefont {Walsworth}}]{PhamPRB2012}%
  \BibitemOpen
  \bibfield  {author} {\bibinfo {author} {\bibfnamefont {L.~M.}\ \bibnamefont
  {Pham}}, \bibinfo {author} {\bibfnamefont {N.}~\bibnamefont {Bar-Gill}},
  \bibinfo {author} {\bibfnamefont {C.}~\bibnamefont {Belthangady}}, \bibinfo
  {author} {\bibfnamefont {D.}~\bibnamefont {Le~Sage}}, \bibinfo {author}
  {\bibfnamefont {P.}~\bibnamefont {Cappellaro}}, \bibinfo {author}
  {\bibfnamefont {M.~D.}\ \bibnamefont {Lukin}}, \bibinfo {author}
  {\bibfnamefont {A.}~\bibnamefont {Yacoby}},\ and\ \bibinfo {author}
  {\bibfnamefont {R.~L.}\ \bibnamefont {Walsworth}},\ }\bibfield  {title}
  {\bibinfo {title} {Enhanced solid-state multispin metrology using dynamical
  decoupling},\ }\href {https://doi.org/10.1103/PhysRevB.86.045214} {\bibfield
  {journal} {\bibinfo  {journal} {Phys. Rev. B}\ }\textbf {\bibinfo {volume}
  {86}},\ \bibinfo {pages} {045214} (\bibinfo {year} {2012})}\BibitemShut
  {NoStop}%
\bibitem [{\citenamefont {Ezzell}\ \emph {et~al.}(2023)\citenamefont {Ezzell},
  \citenamefont {Pokharel}, \citenamefont {Tewala}, \citenamefont {Quiroz},\
  and\ \citenamefont {Lidar}}]{LidarPhysRevAppl2023}%
  \BibitemOpen
  \bibfield  {author} {\bibinfo {author} {\bibfnamefont {N.}~\bibnamefont
  {Ezzell}}, \bibinfo {author} {\bibfnamefont {B.}~\bibnamefont {Pokharel}},
  \bibinfo {author} {\bibfnamefont {L.}~\bibnamefont {Tewala}}, \bibinfo
  {author} {\bibfnamefont {G.}~\bibnamefont {Quiroz}},\ and\ \bibinfo {author}
  {\bibfnamefont {D.~A.}\ \bibnamefont {Lidar}},\ }\bibfield  {title} {\bibinfo
  {title} {Dynamical decoupling for superconducting qubits: A performance
  survey},\ }\href {https://doi.org/10.1103/PhysRevApplied.20.064027}
  {\bibfield  {journal} {\bibinfo  {journal} {Phys. Rev. Appl.}\ }\textbf
  {\bibinfo {volume} {20}},\ \bibinfo {pages} {064027} (\bibinfo {year}
  {2023})}\BibitemShut {NoStop}%
\bibitem [{\citenamefont {Austin}\ \emph {et~al.}(2019)\citenamefont {Austin},
  \citenamefont {Khan}, \citenamefont {Mudassar},\ and\ \citenamefont
  {Chaudhry}}]{c6}%
  \BibitemOpen
  \bibfield  {author} {\bibinfo {author} {\bibfnamefont {S.}~\bibnamefont
  {Austin}}, \bibinfo {author} {\bibfnamefont {M.~Q.}\ \bibnamefont {Khan}},
  \bibinfo {author} {\bibfnamefont {M.}~\bibnamefont {Mudassar}},\ and\
  \bibinfo {author} {\bibfnamefont {A.~Z.}\ \bibnamefont {Chaudhry}},\
  }\bibfield  {title} {\bibinfo {title} {Continuous dynamical decoupling of
  spin chains: Modulating the spin-environment and spin-spin interactions},\
  }\href {https://doi.org/10.1103/PhysRevA.100.022102} {\bibfield  {journal}
  {\bibinfo  {journal} {Phys. Rev. A}\ }\textbf {\bibinfo {volume} {100}},\
  \bibinfo {pages} {022102} (\bibinfo {year} {2019})}\BibitemShut {NoStop}%
\bibitem [{\citenamefont {Soomro}\ and\ \citenamefont
  {Chaudhry}(2022)}]{HudaibaEJPD2022}%
  \BibitemOpen
  \bibfield  {author} {\bibinfo {author} {\bibfnamefont {H.}~\bibnamefont
  {Soomro}}\ and\ \bibinfo {author} {\bibfnamefont {A.~Z.}\ \bibnamefont
  {Chaudhry}},\ }\bibfield  {title} {\bibinfo {title} {Spin chain
  transformations under continuous driving fields},\ }\href@noop {} {\bibfield
  {journal} {\bibinfo  {journal} {Eur. J. Phys. D}\ }\textbf {\bibinfo {volume}
  {76}},\ \bibinfo {pages} {180} (\bibinfo {year} {2022})}\BibitemShut
  {NoStop}%
\bibitem [{\citenamefont {Irfan}\ \emph {et~al.}(2024)\citenamefont {Irfan},
  \citenamefont {Hashmi}, \citenamefont {Zaidi}, \citenamefont {Baig},
  \citenamefont {Ayub},\ and\ \citenamefont {Chaudhry}}]{IrfanPRA2024}%
  \BibitemOpen
  \bibfield  {author} {\bibinfo {author} {\bibfnamefont {A.}~\bibnamefont
  {Irfan}}, \bibinfo {author} {\bibfnamefont {S.~F.~A.}\ \bibnamefont
  {Hashmi}}, \bibinfo {author} {\bibfnamefont {S.~N.}\ \bibnamefont {Zaidi}},
  \bibinfo {author} {\bibfnamefont {M.~U.}\ \bibnamefont {Baig}}, \bibinfo
  {author} {\bibfnamefont {W.}~\bibnamefont {Ayub}},\ and\ \bibinfo {author}
  {\bibfnamefont {A.~Z.}\ \bibnamefont {Chaudhry}},\ }\bibfield  {title}
  {\bibinfo {title} {Continuous dynamical decoupling of spin chains: Inducing
  two-qubit interactions to generate perfect entanglement},\ }\href
  {https://doi.org/10.1103/PhysRevA.109.042622} {\bibfield  {journal} {\bibinfo
   {journal} {Phys. Rev. A}\ }\textbf {\bibinfo {volume} {109}},\ \bibinfo
  {pages} {042622} (\bibinfo {year} {2024})}\BibitemShut {NoStop}%
\bibitem [{\citenamefont {Myerson}\ \emph {et~al.}(2008)\citenamefont
  {Myerson}, \citenamefont {Szwer}, \citenamefont {Webster}, \citenamefont
  {Allcock}, \citenamefont {Curtis}, \citenamefont {Imreh}, \citenamefont
  {Sherman}, \citenamefont {Stacey}, \citenamefont {Steane},\ and\
  \citenamefont {Lucas}}]{MyersonPRL2008}%
  \BibitemOpen
  \bibfield  {author} {\bibinfo {author} {\bibfnamefont {A.~H.}\ \bibnamefont
  {Myerson}}, \bibinfo {author} {\bibfnamefont {D.~J.}\ \bibnamefont {Szwer}},
  \bibinfo {author} {\bibfnamefont {S.~C.}\ \bibnamefont {Webster}}, \bibinfo
  {author} {\bibfnamefont {D.~T.~C.}\ \bibnamefont {Allcock}}, \bibinfo
  {author} {\bibfnamefont {M.~J.}\ \bibnamefont {Curtis}}, \bibinfo {author}
  {\bibfnamefont {G.}~\bibnamefont {Imreh}}, \bibinfo {author} {\bibfnamefont
  {J.~A.}\ \bibnamefont {Sherman}}, \bibinfo {author} {\bibfnamefont {D.~N.}\
  \bibnamefont {Stacey}}, \bibinfo {author} {\bibfnamefont {A.~M.}\
  \bibnamefont {Steane}},\ and\ \bibinfo {author} {\bibfnamefont {D.~M.}\
  \bibnamefont {Lucas}},\ }\bibfield  {title} {\bibinfo {title} {High-fidelity
  readout of trapped-ion qubits},\ }\href
  {https://doi.org/10.1103/PhysRevLett.100.200502} {\bibfield  {journal}
  {\bibinfo  {journal} {Phys. Rev. Lett.}\ }\textbf {\bibinfo {volume} {100}},\
  \bibinfo {pages} {200502} (\bibinfo {year} {2008})}\BibitemShut {NoStop}%
\bibitem [{\citenamefont {Jacobs}(2010)}]{Jacobsbook}%
  \BibitemOpen
  \bibfield  {author} {\bibinfo {author} {\bibfnamefont {K.}~\bibnamefont
  {Jacobs}},\ }\href@noop {} {\emph {\bibinfo {title} {Stochastic Processes for
  Physicists: Understanding Noisy Systems}}}\ (\bibinfo  {publisher} {Cambridge
  University Press},\ \bibinfo {year} {2010})\BibitemShut {NoStop}%
\bibitem [{\citenamefont {Benedetti}\ \emph {et~al.}(2018)\citenamefont
  {Benedetti}, \citenamefont {Salari~Sehdaran}, \citenamefont {Zandi},\ and\
  \citenamefont {Paris}}]{BenedettiPRA2018}%
  \BibitemOpen
  \bibfield  {author} {\bibinfo {author} {\bibfnamefont {C.}~\bibnamefont
  {Benedetti}}, \bibinfo {author} {\bibfnamefont {F.}~\bibnamefont
  {Salari~Sehdaran}}, \bibinfo {author} {\bibfnamefont {M.~H.}\ \bibnamefont
  {Zandi}},\ and\ \bibinfo {author} {\bibfnamefont {M.~G.~A.}\ \bibnamefont
  {Paris}},\ }\bibfield  {title} {\bibinfo {title} {Quantum probes for the
  cutoff frequency of ohmic environments},\ }\href
  {https://doi.org/10.1103/PhysRevA.97.012126} {\bibfield  {journal} {\bibinfo
  {journal} {Phys. Rev. A}\ }\textbf {\bibinfo {volume} {97}},\ \bibinfo
  {pages} {012126} (\bibinfo {year} {2018})}\BibitemShut {NoStop}%
\bibitem [{\citenamefont {Ather}\ and\ \citenamefont
  {Chaudhry}(2021)}]{ChaudhryPRA2021}%
  \BibitemOpen
  \bibfield  {author} {\bibinfo {author} {\bibfnamefont {H.}~\bibnamefont
  {Ather}}\ and\ \bibinfo {author} {\bibfnamefont {A.~Z.}\ \bibnamefont
  {Chaudhry}},\ }\bibfield  {title} {\bibinfo {title} {Improving the estimation
  of environment parameters via initial probe-environment correlations},\
  }\href {https://doi.org/10.1103/PhysRevA.104.012211} {\bibfield  {journal}
  {\bibinfo  {journal} {Phys. Rev. A}\ }\textbf {\bibinfo {volume} {104}},\
  \bibinfo {pages} {012211} (\bibinfo {year} {2021})}\BibitemShut {NoStop}%
\bibitem [{\citenamefont {Mirza}\ and\ \citenamefont
  {Chaudhry}(2024)}]{ChaudhrySciRep2024}%
  \BibitemOpen
  \bibfield  {author} {\bibinfo {author} {\bibfnamefont {A.~R.}\ \bibnamefont
  {Mirza}}\ and\ \bibinfo {author} {\bibfnamefont {A.~Z.}\ \bibnamefont
  {Chaudhry}},\ }\bibfield  {title} {\bibinfo {title} {Improving the estimation
  of environment parameters via a two-qubit scheme},\ }\href@noop {} {\bibfield
   {journal} {\bibinfo  {journal} {Sci. Rep.}\ }\textbf {\bibinfo {volume}
  {14}},\ \bibinfo {pages} {6803} (\bibinfo {year} {2024})}\BibitemShut
  {NoStop}%
\bibitem [{\citenamefont {Mirza}\ and\ \citenamefont
  {Al-Khalili}(2024)}]{mirza2024impact}%
  \BibitemOpen
  \bibfield  {author} {\bibinfo {author} {\bibfnamefont {A.~R.}\ \bibnamefont
  {Mirza}}\ and\ \bibinfo {author} {\bibfnamefont {J.}~\bibnamefont
  {Al-Khalili}},\ }\bibfield  {title} {\bibinfo {title} {The impact of quantum
  correlations on parameter estimation in a spin reservoir},\ }\href@noop {}
  {\bibfield  {journal} {\bibinfo  {journal} {Physica Scripta}\ }\textbf
  {\bibinfo {volume} {99}},\ \bibinfo {pages} {115102} (\bibinfo {year}
  {2024})}\BibitemShut {NoStop}%
\bibitem [{\citenamefont {Zhukas}\ \emph {et~al.}(2021)\citenamefont {Zhukas},
  \citenamefont {Svihra}, \citenamefont {Nomerotski},\ and\ \citenamefont
  {Blinov}}]{ZhukasPRA2021}%
  \BibitemOpen
  \bibfield  {author} {\bibinfo {author} {\bibfnamefont {L.~A.}\ \bibnamefont
  {Zhukas}}, \bibinfo {author} {\bibfnamefont {P.}~\bibnamefont {Svihra}},
  \bibinfo {author} {\bibfnamefont {A.}~\bibnamefont {Nomerotski}},\ and\
  \bibinfo {author} {\bibfnamefont {B.~B.}\ \bibnamefont {Blinov}},\ }\bibfield
   {title} {\bibinfo {title} {High-fidelity simultaneous detection of a
  trapped-ion qubit register},\ }\href
  {https://doi.org/10.1103/PhysRevA.103.062614} {\bibfield  {journal} {\bibinfo
   {journal} {Phys. Rev. A}\ }\textbf {\bibinfo {volume} {103}},\ \bibinfo
  {pages} {062614} (\bibinfo {year} {2021})}\BibitemShut {NoStop}%
\bibitem [{\citenamefont {Abbas}\ \emph {et~al.}(2025)\citenamefont {Abbas},
  \citenamefont {Mirza},\ and\ \citenamefont {Chaudhry}}]{data}%
  \BibitemOpen
  \bibfield  {author} {\bibinfo {author} {\bibfnamefont {R.}~\bibnamefont
  {Abbas}}, \bibinfo {author} {\bibfnamefont {A.~R.}\ \bibnamefont {Mirza}},\
  and\ \bibinfo {author} {\bibfnamefont {A.~Z.}\ \bibnamefont {Chaudhry}},\
  }\href@noop {} {\bibinfo {title} {Data for sensing high-frequency ac fields
  via a two-qubit sensor}},\ \bibinfo {howpublished} {Zenodo, doi: 10.5281/zenodo.17421250} (\bibinfo {year} {2025})\BibitemShut {NoStop}%
\end{thebibliography}
\end{document}